\newcount\Comments
\documentclass[twocolumn]{aastex631}

\usepackage{hyperref}
\usepackage{soul}
\usepackage{xcolor}
\usepackage{graphicx}
\usepackage{amssymb}
\usepackage{array}
\usepackage{multirow}
\usepackage{amsmath}
\usepackage{parskip}
\usepackage{makecell}

\usepackage{bm}

\graphicspath{{./}}
\newcommand{\DELETED}[1]{\relax}%

\definecolor{purple}{rgb}{0.6,0,1}
\definecolor{blue2}{rgb}{0,.5,1}
\newcommand{\kibitz}[2]{\ifnum\Comments=1\textcolor{#1}{#2}\fi}

{\relax}%

\def \HDsnr{14}
\def \HDPILCsnr{12}

\def \SFsnr{1.8}
\def \SFPILCsnr{0.54}

\shorttitle{$K \times \tau$ Forecasts}
\shortauthors{Kramer et al.}

\begin{document}

\title{Cross-correlating the patchy screening and kinetic Sunyaev-Zel'dovich effects as a new probe of reionization}

\author[0000-0003-0238-8806]{Darby M. Kramer}
\affiliation{School of Earth and Space Exploration, Arizona State University, Tempe, AZ 85287, USA}

\author[0000-0002-3495-158X]{Alexander van Engelen}
\affiliation{School of Earth and Space Exploration, Arizona State University, Tempe, AZ 85287, USA}

\author[0000-0001-9420-7384]{Christopher Cain}
\affiliation{School of Earth and Space Exploration, Arizona State University, Tempe, AZ 85287, USA}

\author[0000-0002-8998-3909]{Niall MacCrann}
\affiliation{Department of Applied Mathematics and Theoretical Astrophysics, University of Cambridge, Cambridge CB3 0WA,
United Kingdom}

\author[0000-0001-6778-3861]{Hy Trac}
\affiliation{Department of Physics, Carnegie Mellon University, Pittsburgh, PA 15213}
\affiliation{McWilliams Center for Cosmology and Astrophysics, Carnegie Mellon University, Pittsburgh, PA 15213}

\author[0009-0001-2208-1310]{Skylar Grayson}
\affiliation{School of Earth and Space Exploration, Arizona State University, Tempe, AZ 85287, USA}

\author[0000-0002-3193-1196]{Evan Scannapieco} \affiliation{School of Earth and Space Exploration, Arizona State University, Tempe, AZ 85287, USA}

\author[0000-0002-4495-1356]{Blake Sherwin}
\affiliation{Department of Applied Mathematics and Theoretical Astrophysics, University of Cambridge, Cambridge CB3 0WA,
United Kingdom}

\correspondingauthor{Darby Kramer}
\email{dmkrame1@asu.edu}

\begin{abstract}
The kinetic Sunyaev-Zel’dovich effect (kSZ) and patchy screening effect are two complementary cosmic microwave background (CMB) probes of the reionization era. The kSZ effect is a relatively strong signal, but is difficult to disentangle from other sources of temperature anisotropy, whereas patchy screening is weaker but can be reconstructed using the cleaner polarization channel. Here, we explore the potential of using upcoming CMB surveys to correlate a reconstructed map of patchy screening with (the square of) the kSZ map, and what a detection of this cross-correlation would mean for reionization science. To do this, we use simulations and theory to quantify the contributions to this signal from different redshifts. We then use the expected survey properties for CMB-S4 and CMB-HD to make detection forecasts. We find that, for or our fiducial reionization scenario, CMB-S4 will obtain a hint of this signal at up to $\SFsnr\sigma$, and CMB-HD will detect it at up to $\HDsnr\sigma$. We explore the physical interpretation of the signal and find that it is uniquely sensitive to the first half of reionization and to the bispectrum of the ionized gas distribution.
\end{abstract}


\section{Introduction}\label{sec:intro}

The epoch of reionization (EoR) is a pivotal but elusive time in the history of the Universe. It is the period during which the population of hydrogen  went from being neutral to ionized, likely due to the formation of the first stars and galaxies \citep{scannapieco03,robertson2010,sarmento22}. These first objects are difficult to measure directly because of their faint and highly redshifted nature, and because of the structures that continued to form after the EoR, such as larger, brighter galaxies and galaxy clusters that emit at a variety of wavelengths and intensities. 

Since reionization was caused by individual objects, it is expected to be ``patchy," causing small, ionized bubbles to form and grow over time \citep{gruzinov98,santos03,daloisio15,davies16,puchwein23}. Many experiments and studies seek to extract signals that solely trace the EoR, such as probes of the galaxy population \citep[e.g.,][]{Adams2023}, mean electron optical depth \citep[\textsl{Planck}:][]{planck2018}, neutral hydrogen density as traced by 21cm emission \citep[e.g.][]{Pritchard2012} and Lyman-alpha absorption \citep{Bosman2022} at high redshifts. Each of these probes use unique tools and techniques to access the EoR. 

A particular cosmological tool that can benefit EoR studies in the coming years is the cosmic microwave background (CMB). When the primordial plasma expanded and cooled enough for electrons and protons to combine and form the first atoms at redshift $z \sim 1100$, photons were able to free-stream in all directions as a blackbody background, the CMB, seen today with temperature $T_0 = 2.7$ K. However, there were small density fluctuations traceable by the CMB temperature anisotropies that grew over time to become the large-scale structures we observe today. These temperature fluctuations are known as primary CMB anisotropies and are on the order of 100 $\mu$K today. Furthermore, some of the CMB photons have interacted with the large scale structure that formed in the later Universe, causing secondary CMB anisotropies.

\begin{table*}[htp!]\label{tab:summ_tab}
\hspace{-43pt}
\resizebox{1.07\textwidth}{!}{%
    \begin{tabular}{|c|c|c|c|}
        \hline
        \textbf{Spectrum Type} & \textbf{Reionization Probe} & \textbf{Example Reference(s)} & \textbf{Forecasted ``S4'' SNR Range} \\
        \Xhline{2\arrayrulewidth}
        \multirow{4}{*}{Auto-spectra} & $\langle T_\mathrm{kSZ} \times T_\mathrm{kSZ}\rangle$ & \citet{Raghunathan2023} & $\sim 70-80\sigma$ \\
        \cline{2-4}
        & $\langle K \times K \rangle$ & \citet{SmithFerraro2017} & $\sim 100\sigma$ \\ 
        \cline{2-4}
        & $\langle \tau \times \tau \rangle$ & \citet{dvorkinsmith2009}, \citet{Jain2024} & $\sim 2-3\sigma$\\ 
        \cline{2-4}
        & $\langle y \times y \rangle$ & \citet{sun2024} & $\sim 3\sigma$ \\
        \Xhline{4\arrayrulewidth}
        \multirow{5}{*}{Cross-spectra} & $\langle \tau \times T_{21} \rangle$ & \citet{Meerburg2013} & $\sim 10\sigma$ \\
        \cline{2-4}
        & $\langle K \times T_{21} \rangle$ & \citet{Ma2018} & $\sim 50\sigma$ \\
        \cline{2-4}
        & $\langle \tau \times \kappa_{\mathrm{CMB}} \rangle$ & \citet{FengHolder2019}, \citet{Bianchini2023} & $\sim 10-22\sigma$ \\
        \cline{2-4}
        & $\langle \tau \times y \rangle$ & \citet{namikawa2021}, \citet{remazeilles2024} & $\sim 2-7\sigma$ \\
        \cline{2-4}
        & $\langle K \times \delta_{\mathrm{g}} \rangle$ & \citet{LaPlante2022} & $\sim 4-11\sigma$ \\
        \cline{2-4}
        & $\langle y \times \delta_{\mathrm{g}} \rangle$ & \citet{baxter2021} & $< 2\sigma$ \\
        \Xhline{2\arrayrulewidth}
    \end{tabular}%
}

    \caption{Review of some previously-forecasted probes of reionization with CMB data and their reported signal-to-noise ratios. Above the thick horizontal line are the three auto-power spectra discussed in Section \ref{sec:intro} that contain reionization information. Below this line are a few cross-correlations that have been studied in the literature attempting to isolate the reionization information in their constituents. The CMB-S4 (or S4-like) signal-to-noise forecasts for these probes are in the rightmost column.}
\end{table*}

\begin{figure*}
    \centering
    \includegraphics[width=\textwidth]{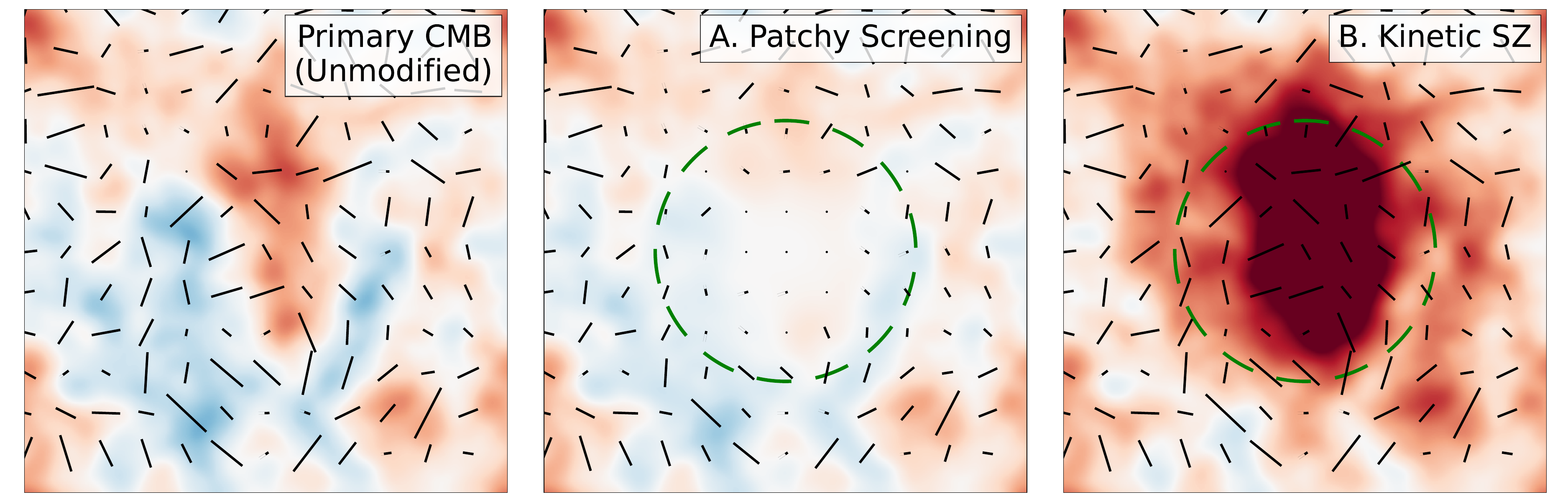}
    \caption{A demonstration of the map-level effects of patchy screening and the kSZ effect. The left panel shows the primary, unmodified CMB. Panel A (center) demonstrates the screening that happens to the primary CMB when a bubble of electrons is simulated in the foreground, located inside the dashed green circle. Note that the CMB fluctuations, both temperature (colors) and polarization (black vectors), are "washed out" by the Thomson scattering. Panel B (right) demonstrates the shift in the CMB temperature fluctuations that happens when the bubble of electrons is given a positive LOS velocity with respect to the Hubble flow. This leads to a colder (redder) CMB spot in temperature; the polarization is unaffected at this order in scattering.}
    \label{fig:threepanel}
\end{figure*}

A set of these secondary interactions occurs when the CMB photons Thomson scatter off bubbles of free electrons, an effect called patchy screening \citep[e.g., ][]{zahn2007, dvorkinsmith2009, natarajan2013}. This scattering acts as a screen on the CMB with some optical depth $\mathbf{\tau}$, washing out the primary signal and generating new polarization \citep{zaldarriaga1997} on large angular scales (see Fig. \ref{fig:threepanel}). 
Detections of the all-sky average value of $\tau$ have been made using large-scale polarization with both \textsl{WMAP} \citep[$\bar{\tau} = 0.087 \pm 0.017$,][]{WMAP2009} and \textsl{Planck} \citep[$\bar{\tau} = 0.054 \pm 0.007$,][]{planck2018}. The angular fluctuations in the $\tau$ field, $\tau(\hat{\mathbf{n}})$, reduce the CMB signal more strongly along lines of sight where the electrons are more dense. Both high-redshift reionization processes ($z\gtrsim6$) and low-redshift galaxies ($z\lesssim3$) create free electrons that contribute to this signal. The EoR $\tau$ fluctuations, ``patchy $\tau$," are expected to have an order of magnitude larger variance than their post-EoR counterpart \citep[e.g., Fig. 2 in][]{roy2022probing}. Even so, $\tau(\hat{\mathbf{n}})$ for all redshifts is more difficult to detect than its average value due to current CMB survey noise levels. 

Forecasts estimate that the patchy $\tau$ auto-power spectrum will likely be detectable with future CMB surveys \citep{dvorkinsmith2009,Jain2024}. \citet{roy2022probing} forecasted a cross-correlation of patchy $\tau$ with galaxies at post-EoR redshifts to constrain the circumgalactic medium (CGM) around these galaxies, and there is a hint of $\tau(\hat{\mathbf{n}})$ using this method by \citet{coulton2025} with CMB data from the Atacama Cosmology Telescope (ACT) and \textsl{Planck} in cross-correlation with galaxy data from $0<z<1$. Other works have forecasted cross-correlations between patchy $\tau$ and other tracers of the reionization process to increase the signal-to-noise on EoR measurements \citep{Meerburg2013,FengHolder2019,Bianchini2023,namikawa2021}, as well as second helium reionization at lower redshifts \citep{caliskan2023}, but thus far no other detections of these fluctuations have been made.

A similar secondary CMB anisotropy, the kinetic Sunyaev-Zeldovich (kSZ) effect, causes inverse-Compton scattering of the CMB, shifting the observed temperatures of the photons depending on the electrons' line-of-sight (LOS) velocity relative to the Hubble flow while preserving their blackbody distribution (see Fig. \ref{fig:threepanel}). As with patchy screening, the kSZ effect traces electron fluctuations from both reionization and from the CGM surrounding lower-redshift galaxies. The first statistical measure of the kSZ signal was found in data from the Atacama Cosmology Telescope (ACT) and the Baryon Oscillation Spectroscopic Survey by \citet{Hand2012}, while evidence specifically for the auto-power spectrum was seen at $\sim3\sigma$ by the South Pole Telescope team \citep[SPT,][]{Reichardt_2021}, and is forecasted at up to $80\sigma$ \citep{Raghunathan2023} by the CMB-Stage 4 survey \citep[CMB-S4,][]{Abazajian2019,Abazajian2022}. However, the power spectra of the high- \citep[e.g.,][]{Trac2011} and low- \citep[e.g.,][]{Shaw2012} redshift kSZ components have comparable amplitudes and shapes, so their effects are practically indistinguishable without knowing one of the components \citep[see, e.g., Fig. 1 in the arXiv version of][]{SmithFerraro2017}. Thus, an unambiguous measurement of the kSZ effect solely from the EoR remains elusive. 

The kSZ effect has an expectation value of zero across the sky because it traces the peculiar velocity field, and there are no preferred directions in the Universe. Therefore, in order to perform joint analyses with the kSZ effect on the full sky, it is necessary to use higher-order kSZ functions. Many studies have weighted the kSZ halos by their reconstructed LOS velocities before stacking \citep[e.g.,][]{schaan2021, Hadzhiyska2024}, reconstructed velocities given the CMB map and a galaxy catalog \citep[e.g., ][]{mccarthy2024, lague2024}, or performed pairwise stacking \citep[e.g., ][]{Hand2012, gallardo2021}; the connection between these methods is demonstrated by \citet{smith2018}. A contrasting method to these is to perform three-point correlations with two legs in the kSZ signal and one in another tracer \citep[``projected-fields,"][]{Dore2004, DeDeo2005, Hill2016, FerraroHill2016, Kusiak2021, bolliet2023}. This effectively calculates the squared kSZ field with a reconstruction process similar to that of CMB lensing \citep{hu2001,huokamoto2002}, and prepares the kSZ signal to be cross-correlated with another tracer of the matter density fluctuations. 

Alternatively, given the squared kSZ map, $K$, one can compute the autospectrum, $C_L^{KK}$ to probe the electron overdensities. This spectrum is a measure of the kSZ trispectrum, or four-point function, which yields a measurement sensitive to velocity fluctuations on large scales \citep{SmithFerraro2017, FerraroSmith2018} and electron density fluctuations on small scales. Two recent studies, \citet{raghunathan2024} and \citet{maccrann2024}, placed upper limits on this autospectrum using South Pole Telescope (SPT) + Herschel-SPIRE and ACT + \textsl{Planck} CMB data, respectively. The CMB temperature field contains a large amount of non-Gaussian, extragalactic foregrounds, such as from the cosmic infrared background (CIB) and the thermal Sunyaev-Zel'dovich (tSZ) effect. There are methods to mitigate these foregrounds, such as combining multi-frequency data via internal linear combinations (ILC) \citep[e.g., ][]{remazeilles2011,Raghunathan2023}, or with aggressive masking, but they come at the cost of signal-to-noise reduction. Furthermore, these methods leave behind some foreground trispectrum contamination; both \citet{raghunathan2024} and \citet{maccrann2024} found that while using ILC methods the temperature trispectra from foregrounds such as the tSZ and CIB are still the dominant contributions to this auto-power spectrum, making it more difficult to probe the EoR with $C_L^{KK}$ than anticipated in \citet{SmithFerraro2017}.

Table \ref{tab:summ_tab} summarizes how well CMB-S4 may measure several auto- and cross-correlations with observables that could probe EoR science. The first four are auto-correlations studied in \citet{Raghunathan2023}, \citet{SmithFerraro2017}, \citet{dvorkinsmith2009,Jain2024}, and \citet{sun2024}. These auto-power spectra contain important information about the EoR, such as its midpoint and duration. However, low-redshift free electrons in the CGM also contribute a significant amplitude to each, making it difficult to just probe the EoR with them. The remaining references in Table \ref{tab:summ_tab} are studies that seek to isolate the reionization portions of these signals using cross-correlations of different CMB EoR probes, including $K$ and $\tau$. These show some promise for detections with CMB-S4, as shown in the forecasts. 

The distribution of neutral gas, as traced by the brightness temperature of 21 cm photons emitted from the neutral hydrogen in the Universe, $T_{21}$, has an appreciable anti-correlation with $\tau$ because the ionized and neutral regions trace opposite sides of the bubble boundaries. As shown in Table \ref{tab:summ_tab}, the Square Kilometre Array (SKA) $T_{21}$ $\times$ CMB-S4 $\tau$ data might have the statistical power to make a first detection, but $T_{21}$ foreground mitigation makes this a challenging prospect due to the filtering that must be applied \citep{Meerburg2013}. $T_{21}$ is expected to have a larger signal-to-noise when cross-correlated with $K$ for roughly the same experiments \citep{Ma2018}, but has the same filtering issue. One could also potentially reconstruct the remote dipoles during reionization using the same datasets \citep{Hotlini2022}. There is also promise of a detection of the cross-correlation between CMB lensing ($\kappa_{\mathrm{CMB}}$) with $\tau$ at high significance, where both maps are reconstructed with a survey like CMB-S4 \citep{FengHolder2019,Bianchini2023}. Presumably, many reionization models will be ruled out due to the high signal-to-noise expected on this cross-correlation. 

A cross-correlation between CMB-S4 patchy $\tau$ and the Compton $y$ parameter, a measure of the pressure profiles in bubbles from the tSZ effect, has been forecasted at $\sim 7\sigma$ by \citet{namikawa2021}, but only 1.6$\sigma$ by \citet{remazeilles2024}. Depending on the physics of reionization, these forecasts show that $C_L^{\tau y}$ could be an informative probe of reionization parameters. Another cross-correlation that probes reionization physics is one between overdensities of high redshift galaxies ($\delta_g$) and the $K$ field, since galaxies were likely the driving force behind reionization. \citet{LaPlante2022} find that, even while accounting for foreground power, an appreciable detection of this $K \times \delta_g$ signal may be made with future surveys such as CMB-S4 and Roman, despite the relatively low number densities for galaxies found at these high redshifts. Finally, the study performed in \citet{baxter2021} forecasts that a cross-correlation between high-redshift galaxy overdensities and Compton $\mathbf{y}$ will yield information about reionization, but only at low signal-to-noise due to low-redshift tSZ and instrumental noise.

We note that correlations between $K$ and $\tau$ have not been considered in the literature before, and both will be available with the same set of planned CMB surveys, needing no ancillary datasets. We expect that this cross-correlation will be a positive signal due to the fact that the constituents probe the same bubbles: A given bubble will appear as a positive fluctuation in both the electron optical depth and the square of the kSZ. Furthermore, measuring this signal could have several advantages over measuring either signal on its own due to their individual weaknesses. The large impact of temperature foregrounds, both in residual $TT$ foreground power and in the trispectrum, on a measurement of $C_L^{KK}$ can be significantly reduced by measuring $\tau$ with the polarization-based estimator and cross-correlating, since the CMB foregrounds are weaker in polarization than in temperature. This polarizatioxn-based estimator is what we use for forecasting in this work. Furthermore, cross correlating $\tau$ with $K$ should increase the signal-to-noise when compared to forecasts of the $\tau$ autospectrum because $K$ is projected to be measured by CMB-S4 at $\sim 100\sigma$ \citep{SmithFerraro2017}. This cross-correlation of the squared kSZ field with the $\tau$ field at high redshifts is the analog of the projected-fields kSZ methods of, e.g., \citet{Dore2004} and \citet{Hill2016}, which focus on the post-EoR kSZ. 

In this paper, we assess the possibility of measuring the cross-correlation between the $\tau$ and $K$ fields in the context of probing the EoR. This prospective cross-correlation signal remains unexplored, and it could yield a new measurement of EoR parameters. To explore this new estimator, in Section \ref{sec:formalism} we lay out the formalism of the $\tau$ and $K$ signals from the data and explore the aspects of the EoR that a cross-correlation between these two fields would constrain. We then discuss the reconstruction of these signals (Section \ref{sec:reconstruction}), the expected signals according to reionization simulations (Section \ref{sec:ambermocks}), and their associated noise (Section \ref{sec:noise}). We also use post-EoR simulations to quantify the low-redshift contribution to this signal in Section \ref{sec:agora}. Section \ref{sec:forecasts} contains our signal-to-noise forecasts for detecting this cross-correlation with upcoming surveys, and how sensitive this signal is to EoR parameters. We validate and interpret these results in Section \ref{sec:theoryint} and discuss the results and outlook in Section \ref{sec:discussion}. 

\section{Formalism}\label{sec:formalism}

In this section, we define the fields $K$ and $\tau$ in real space before taking their cross-power spectrum and exploring its implications analytically. Note that the derivations in this work all assume the flat-sky approximation for simplicity, though in the analysis we use the full curved-sky treatment with spherical harmonics. This formalism closely follows that of \citet{Dore2004}, which introduced the ideas of projected-field kSZ measurements. 

The observed CMB temperature $T(\hat{\mathbf{n}})$ along line of sight $\hat{\mathbf{n}}$ contains information from many secondary anisotropic effects. The primary, unmodified CMB temperature $\tilde{T}$ is observed through a potentially anisotropic optical depth $\tau$, and also contains an additive term from the kSZ effect:
\begin{equation}
    T(\hat{\mathbf{n}}) \supset \tilde{T}(\hat{\mathbf{n}})\ e^{-\tau(\hat{\mathbf{n}})} + \Delta T_{\mathrm{kSZ}}(\hat{\mathbf{n}}).
\end{equation}
Here the tilde represents the primary CMB temperature. We have not included other scattering terms, including the other flavors of the SZ effect, the impact of gravitational lensing, or other emissive sources like radio or dusty galaxies. 

\subsection{$\tau$}
The fluctuations in the Thomson optical depth of the Universe can be described as the integral of the visibility function and the electron density fluctuations over time:
\begin{equation}\label{eq:tau_n}
    \tau(\hat{\mathbf{n}}) = \int_0^{\eta_{\mathrm{CMB}}} d\eta\ g(\eta)\ \left[1 + \delta_e(\mathbf{r},\eta)\right],
\end{equation}
where the electron overdensity $\delta_e(\mathbf{r}, \eta)$ at comoving position $\mathbf{r}$ and conformal lookback time $\eta$ is given in terms of both the gas overdensity and the ionization state \citep[as in, e.g., ][]{alvarez2016}. $g(\eta)$, the visibility function, or the probability that an observed photon last-scattered into our line of sight at look-back time $\eta$, is given by
\begin{equation}\label{eq:visfunc}
    g(\eta) = \sigma_T\ \bar{n}_e(\eta)\ a(\eta)\ e^{-\tau(<\eta)}.
\end{equation}
Here, $\sigma_T = \frac{8\pi}{3}(\frac{e^2}{m_e c^2})^2 = 6.7 \times 10^{-25} \text{cm}^2$ is the Thomson scattering cross section, $a(\eta)$ is the scale factor, $\bar{n}_e(\eta)$ is the spatially-averaged number density of electrons in the Universe at a given conformal look-back time, and $\tau(<\eta)$ is the optical depth up to $\eta$. 

Patchy screening affects both the CMB temperature ($T$, proportional to the intensity) and polarization (Stokes $Q$ and $U$ parameters) fields in the following ways:

\begin{equation}\label{eq:taun_temp}
    T(\hat{\mathbf{n}}) = e^{-\tau(\hat{\mathbf{n}})} \tilde{T}(\hat{\mathbf{n}}),
\end{equation}
\begin{equation}\label{eq:taun_pol}
    (Q \pm iU)(\hat{\mathbf{n}}) = e^{-\tau(\hat{\mathbf{n}})}(\tilde{Q} \pm i\tilde{U})(\hat{\mathbf{n}}).
\end{equation}

The $\tau$ field is calculable as a quadratic estimator of CMB fields, whether from temperature, polarization, or both. This method was originally developed for measuring CMB lensing: lensing couples different scales of the CMB fluctuations, turning the CMB into a statistically anisotropic field as we see it. Patchy $\tau$ has a similar coupling effect, so it therefore can be measured with a similar quadratic estimator \citep{dvorkinsmith2009}. We define this estimator in Section \ref{sec:reconstruction}.

\subsection{kSZ$^2$, or K}

The dimensionless kSZ field, $\Theta_{\mathrm{kSZ}}(\hat{\mathbf{n}}) = \Delta T_\mathrm{kSZ}(\hat{\mathbf{n}})/T_0$, traces the momentum field such that
\begin{equation} \label{eq:ksz_n}
    \Theta_\mathrm{kSZ}(\hat{\mathbf{n}}) = - \int_0^{\eta_\mathrm{CMB}} d\eta \, g(\eta) \, \mathbf{p}_e \cdot \hat{\mathbf{n}},
\end{equation}
or, with more resemblance to Eq. \ref{eq:taun_temp},
\begin{equation} \label{eq:ksz_n_2}
    \Theta_\mathrm{kSZ}(\hat{\mathbf{n}}) = - \int_0^{\eta_\mathrm{CMB}} d\eta \, g(\eta) \, [1+\delta_e(\mathbf{r}, \eta)]\mathbf{v}_e \cdot \hat{\mathbf{n}}
\end{equation}
with the visibility function as in Eq. \ref{eq:visfunc}. Note that we define $\hat{\mathbf{n}}$ such that it is directed away from the observer toward the sky, and therefore the minus sign indicates a positive kSZ fluctuation when the peculiar velocity points toward the observer.

As mentioned before, the $K$ field is defined using the square of the $\Theta_{\mathrm{kSZ}}$ field:
\begin{equation}
    K(\hat{\mathbf{n}}) \sim \Theta_\mathrm{kSZ}^{\mathrm{filt}}(\hat{\mathbf{n}})^2
    \label{eq:filtksz}
\end{equation}
where $\Theta_{\text{kSZ}}^{\mathrm{filt}}(\hat{\mathbf{n}})$ is a filtered kSZ map and the tilde represents that it is not a pure proportionality due to a final renormalization step. The filter on the kSZ map (defined in Section \ref{sec:reconstruction}, shown in Fig. \ref{fig:filters}) serves to isolate the scales where the fluctuations in the map have significant contribution from $\delta_e$ weighted by their velocities. In Section \ref{sec:reconstruction}, we precisely define this $K$ field as a quadratic estimator of the CMB temperature field. 

\subsection{$K \times \tau$}

To understand what a cross-correlation between $K$ and $\tau$ probes in an analytical way, we closely follow the reasoning in \citet{Dore2004}, \citet{DeDeo2005}, and \citet{FerraroHill2016}. These studies discuss the cross-correlation between the kSZ$^2$ signal ($K$) and other tracers of large scale structure: cosmic shear and galaxy overdensities, generally at much lower redshifts than the EoR signal we focus on here. However, their broad assumption that velocity and matter fluctuations dominate the $K$ signal at different scales applies in the same manner.

Following Eq. \ref{eq:tau_n}, $\tau(\hat{\mathbf{n}})$ is a map of the ionized electrons projected along the line of sight with a redshift kernel given by the CMB visibility function. $\Theta_{\text{kSZ}}(\hat{\mathbf{n}})$ is, per Eq. \ref{eq:ksz_n_2}, a similarly projected map of the same ionized electrons, only additionally weighted by the LOS component of their velocities. We can thus schematically denote $\tau(\hat{\mathbf{n}}) \sim \int \delta_e$ and $\Theta_{\text{kSZ}}(\hat{\mathbf{n}}) \sim \int v_r \delta_e$, where the $\int$ symbol refers to the LOS projection. The $\tau$ autospectrum therefore measures the amplitude of the fluctuations in the electron density, $C_L^{\tau\tau} \sim \int \langle \delta_e \delta_e \rangle$, and the $\Theta_{\text{kSZ}}$ autospectrum measures those same fluctuations convolved with the fluctuations in the electron velocities, $C_l^{TT,\mathrm{kSZ}} \sim \int \langle v_r \delta_e v_r \delta_e \rangle$ \citep[e.g., ][]{Ma2002}. Given that $K(\hat{\mathbf{n}}) \sim \int v_r^2 \delta_e^2$, the $K$ autospectrum probes the four-point function of the kSZ, $C_L^{KK} \sim \int \langle v_r^4 \delta_e^4 \rangle$ \citep{SmithFerraro2017}. It therefore follows that a cross-correlation between the $K$ and $\tau$ fields is a five-point function, $C_L^{K\tau} \sim \int \langle \delta_e v_r \delta_e v_r \delta_e \rangle$.

On scales of several arcminutes we expect $\delta_e$ to vary significantly but the velocities to be coherent \citep[e.g.,][]{Ma2002}. It is thus reasonable to ask whether these terms will decouple; if so, the $C_L^{K\tau}$ statistic will trace $\left< v_r v_r \right> \left< \delta_e \delta_e \delta_e \right>$, yielding a measure of the electron bispectrum multiplied by the well-understood statistics of the large-scale cosmic bulk flows. This is the same argument that \citet{Dore2004} make and several ``projected-fields" studies thereafter follow \citep{Hill2016,FerraroHill2016,Kusiak2021,bolliet2023}. The differences here compared to the projected-fields studies are that we use the electron overdensities instead of galaxy ones, and we are probing EoR redshifts rather than post-EoR redshifts. We evaluate this further in Section \ref{subsec:bispectest}. 

A theoretical expression for $C_L^{K\tau}$, valid in all regimes, is
\begin{equation}\label{eq:clkt}
    C_L^{K\tau}\ = A^K_{TT}(L) \int_0^{\eta_{\mathrm{CMB}}} \frac{d\eta}{\eta^2}\ g^3(\eta)\ \mathcal{T}\left(k=\frac{L}{\eta}, \eta\right)
\end{equation}
where $A^K_{TT}(L)$ is a normalization factor that we apply to our reconstructed $K$ maps in this work and define in the next section. Because $\tau$ and $\Theta_\mathrm{kSZ}$ are sourced by the same free electrons, the cross-correlation between the $\tau$ and $K$ maps is sensitive to $g^3(\eta)$; for a more general tracer with radial distribution $W(\eta)$, this factor would instead be $W(\eta)g^2(\eta)$, \citep[e.g., ][]{Dore2004}. $\mathcal{T}\left(k, \eta\right)$ is defined as the triangle power spectrum:
\begin{multline}
    \mathcal{T}(k, \eta) \equiv \\
    \int \frac{d^2 \mathbf{q}}{(2\pi)^2}\ W^\mathrm{kSZ}(\mathbf{q}\eta)\ W^\mathrm{kSZ}(|\mathbf{k} + \mathbf{q}|\eta)\ B_{\delta_e p_{\hat{\mathbf{n}}} p_{\hat{\mathbf{n}}}} (\mathbf{k}, \mathbf{q}, -\mathbf{k} - \mathbf{q}),
    \label{eq:triangle}
\end{multline}
where $W^\mathrm{kSZ}$ are the filters referenced within Equation \ref{eq:filtksz} and shown in Figure \ref{fig:filters} below. $B_{\delta_e p_{\hat{\mathbf{n}}} p_{\hat{\mathbf{n}}}} (\mathbf{k}, \mathbf{q}, -\mathbf{k} - \mathbf{q})$ is the ``hybrid bispectrum" of the electron density fluctuations $\delta_e$ and the LOS component of the electron momentum field $p_{\hat{\mathbf{n}}}$:

\begin{align}
    \langle \delta_e(\mathbf{k}_1)\ p_{\hat{\mathbf{n}}}(\mathbf{k}_2)\ p_{\hat{\mathbf{n}}}(\mathbf{k}_3) \rangle 
    &= \nonumber \\
    (2\pi)^3 \delta_D(\mathbf{k}_1+\mathbf{k}_2+\mathbf{k}_3) & \ B_{\delta_e p_{\hat{\mathbf{n}}} p_{\hat{\mathbf{n}}}}(\mathbf{k}_1,\mathbf{k}_2,\mathbf{k}_3).
    \label{eq:bispecdef}
\end{align}

The triangle power spectrum arises because taking the square of the kSZ signal at the map level leads to a convolution in the Fourier domain: It represents a compression of the hybrid bispectrum $B_{\delta_e p_{\hat{\mathbf{n}}} p_{\hat{\mathbf{n}}}}$ as shown by the choice of triangles with one side-length equal to $k$.

Therefore, this cross-correlation of the $K$ and $\tau$ fields exists and should be non-zero. Furthermore, we expect it to be positive because the constituents trace common electron overdensities. Finally, we predict that this cross-correlation traces both EoR and post-EoR physics, just as the patchy $\tau$ autospectrum and the $K$ autospectrum do \citep{roy2022probing, SmithFerraro2017}. We quantify the high- and low-redshift components of $C_L^{K\tau}$ using simulations described in Sections \ref{sec:ambermocks} and \ref{sec:agora}, respectively.

\section{Field Reconstruction}\label{sec:reconstruction}
\begin{table*}
    \centering
    \begin{tabular}{|c|c|c|c|}
         \hline
         $XY$ & $f^{\tau}_{XY}(\mathbf{l}_1,\mathbf{l}_2)$& $f^{K}_{XY}(\mathbf{l}_1,\mathbf{l}_2)$& $f^{\phi}_{XY}(\mathbf{l}_1,\mathbf{l}_2)$\\[0.5ex] 
         \Xhline{2\arrayrulewidth}
         $TT$ & $C_{l_1}^{TT}+C_{l_2}^{TT}$ & $\sqrt{C_{l_1}^{TT,\mathrm{kSZ}}C_{l_2}^{TT,\mathrm{kSZ}}}$ 
         & $(\mathbf{L} \cdot \mathbf{l}_1)C_{l_1}^{TT}+(\mathbf{L} \cdot \mathbf{l}_2)C_{l_2}^{TT}$ \\
         \hline
         $EB$ & $[\tilde{C}_{l_1}^{EE} - \tilde{C}_{l_2}^{BB}] \sin [2\varphi_{l_1 l_2}]$ & ---& 
         $[(\mathbf{L} \cdot \mathbf{l}_1)\tilde{C}_{l_1}^{EE} - (\mathbf{L} \cdot \mathbf{l}_2)\tilde{C}_{l_2}^{BB}]\sin [2\varphi_{l_1 l_2}]$\\
         \hline
    \end{tabular}
    
    \centering
        \caption{The mode-coupling expressions that enter the minimum variance filters for $\hat{\tau}$, $\hat{K}$, and $\hat{\phi}$ given CMB signals $X$ and $Y$, where $\varphi_{l_1} = \cos^{-1}(\hat{\mathbf{n}}\cdot\hat{l_1})$, $\varphi_{l_1 l_2} = \varphi_{l_1} - \varphi_{l_2}$, and $\mathbf{L} = \mathbf{l}_1+\mathbf{l}_2 $. In this work, $C_{l}^{TT,\mathrm{kSZ}}$ is the fiducial theory kSZ power spectrum from AMBER \citep{AMBER2022}.}  \label{tab:filters}
\end{table*}
In practice, both $\tau$ and $K$ are reconstructed from CMB data using quadratic estimators, similar to those used routinely for lensing reconstruction. Obtaining maps of $\tau$ and $K$ in order to cross-correlate them involves isolating specific modes of the CMB temperature and polarization data and multiplying them together in the map domain. This process up-weights the scales that are dominated by the signals of interest. In this section, we introduce the flat-sky approximations of the quadratic estimators for the $K$ and $\tau$ fields in a general form before defining their different filters \citep[as done in, e.g.,][]{suyadav2011}. These general equations are the same as those for the CMB lensing potential, $\phi$, quadratic estimators \citep{huokamoto2002}, whose filters we also present in Table \ref{tab:filters} for reference.

The cross-power spectrum of the Fourier transform of two CMB fields $X$ and $Y$ can be expressed as
\begin{equation}
     \langle X(\mathbf{l}_1)Y(\mathbf{l}_2)\rangle_\mathrm{CMB} = (2\pi)^2\ \delta(\mathbf{l}_1 + \mathbf{l}_2)\ C_{l}^{XY},
\end{equation}
where $\mathbf{l}_1$ and $\mathbf{l}_2$ are the wavevectors describing the plane waves on the sky on our flat surface. The CMB subscript indicates an ensemble average over multiple realizations of the CMB.

Given observed CMB maps $X$ and $Y$, representing temperature or polarization maps $T$, $E$, or $B$, a quadratic estimator for a given field $\Gamma$ which couples modes $\mathbf{l}_1$ and $\mathbf{l}_2\equiv\mathbf{L}-\mathbf{l}_1$ can be written as 
\begin{equation}
    \begin{split}
    & \hat{\Gamma}_{XY}(\mathbf{L}) = \\
    & A^{\Gamma}_{XY}(\mathbf{L}) \int \frac{d^2 \mathbf{l}_1}{(2\pi)^2} F^{\Gamma}_{XY}(\mathbf{l}_1, \mathbf{L}-\mathbf{l}_1) \left[X(\mathbf{l}_1)Y(\mathbf{L}-\mathbf{l}_1)\right],
    \end{split} \label{eq:genrecon}
\end{equation}
where $A^{\Gamma}_{XY}(\mathbf{L})$ is the normalization function,  
\begin{equation} 
    A_{XY}^{\Gamma}(\mathbf{L}) = \left[ \int \frac{d^2 \mathbf{l}_1}{(2\pi)^2} F^{\Gamma}_{XY}(\mathbf{l}_1, \mathbf{L}-\mathbf{l}_1) f^{\Gamma}_{XY}(\mathbf{l}_1, \mathbf{L}-\mathbf{l}_1) \right]^{-1} \label{eq:gennorm},
\end{equation}
and $F^{\Gamma}_{XY}(\mathbf{l}_1, \mathbf{L}-\mathbf{l}_1)$ is the filtering function,
\begin{equation}
    F^{\Gamma}_{XY}(\mathbf{l}_1, \mathbf{L}-\mathbf{l}_1) = \frac{f^{\Gamma}_{XY}(\mathbf{l}_1, \mathbf{L}-\mathbf{l}_1)}{\alpha C_{l_1}^{XX,\mathrm{t}} C_{l_2}^{YY,\mathrm{t}}} \label{eq:gencapf},
\end{equation}
with $C_{l}^{XX,\mathrm{t}} = C_{l}^{XX} + C_{l}^{XX,\ \text{noise}}$, the power spectrum of the total sky emission, including lensed CMB, foregrounds, and instrumental noise. $\alpha=1$ or $2$ depending on the symmetry properties of the estimator; for $XY = TT$, $EE$, or $BB$, $\alpha=2$ \citep{huokamoto2002, suyadav2011}. $A^{\Gamma}_{XY}(\mathbf{L})$ normalizes the estimate $\hat{\Gamma}$ such that it is, on average, equal to the actual $\Gamma$ field. The mode-coupling functions $f^{\Gamma}_{XY}(\mathbf{l}_1,\mathbf{L})$ are given in Table \ref{tab:filters} for the $\tau$, $K$, and $\phi$ mode coupling choices. When assuming ``ideal" filtering, meaning that the total power spectra in the filter match the total power in the maps, the noise power spectrum for a given mode-coupling becomes equal to the normalization; i.e., $N^{\Gamma}_{XY}(\mathbf{L}) = A^{\Gamma}_{XY}(\mathbf{L})$.

\subsection{$\tau$ Reconstruction \& Normalization}\label{subsec:taunorm}
Patchy optical depth $\tau$ modulates CMB maps according to Equations \ref{eq:taun_temp} and \ref{eq:taun_pol}, causing new correlations between CMB modes $\mathbf{l}_1$ and $\mathbf{l}_2$ according to:
\begin{equation}
     \langle X(\mathbf{l}_1) Y(\mathbf{l}_2)\rangle_\mathrm{CMB}\ = f^{\tau}_{XY}(\mathbf{l}_1, \mathbf{l}_2)\ \tau(\mathbf{l}_1+\mathbf{l}_2)\label{eq:taumodl},
 \end{equation}
with an assumed-fixed field $\tau$ and $f^{\tau}_{XY}(\mathbf{l}_1,\mathbf{l}_2)$ given in Table \ref{tab:filters}. See the bottom two panels in Fig. \ref{fig:filters} for the specific $E$ and $B$ filters used here for $\tau$. One can reconstruct this patchy $\tau$ signal in Fourier space, $\hat{\tau}(\mathbf{L})$, at a scale $\mathbf{L}\neq 0$ using the quadratic estimator derived in \citet{dvorkinsmith2009}, whose flat sky representation is presented in \citet{suyadav2011} This has been applied to real data in e.g. \citet{verascreening2013} and \citet{toshiyatau2018}, and is equivalent to Equation \ref{eq:genrecon} with $\Gamma = \tau$.

\subsection{K Reconstruction \& Normalization} \label{subsec:Krecon}
The $K(\mathbf{L})$ field is reconstructed in a similar manner to $\tau(\mathbf{L})$ and $\phi(\mathbf{L})$, by way of a quadratic estimator which filters and multiplies two CMB maps together. In this case, $K(\mathbf{L})$ is only measurable with a quadratic estimator of the CMB temperature field. This is because the kSZ effect only traces the electron bulk flows, and is not dependent on the incident photons' polarization. Reconstructed in the manner laid out in Section \ref{sec:reconstruction}, $\hat{K}(\mathbf{L})$ traces fluctuations in the (squared) velocity field on large scales \citep{SmithFerraro2017} but is equivalent, up to potentially the final normalization step, to the squared kSZ fields constructed for projected-fields studies on smaller scales \citep{Dore2004, Hill2016}. We use the same symbol $K$ to refer to both regimes.

Table \ref{tab:filters} contains the minimum variance filter for $\hat{K}(\mathbf{L})$ reconstruction according to equations \ref{eq:genrecon}, \ref{eq:gennorm}, and \ref{eq:gencapf} for $\Gamma = K$. See the second panel in Fig. \ref{fig:filters} for the filters used to reconstruct the $K$ signal in this work. We discuss the origin of the input kSZ signal to this reconstruction in the next section, and discuss our methods for computing the optimal reconstruction noise (as in Eq. \ref{eq:genrecon}) for CMB-S4 and CMB-HD in Section \ref{sec:noise}.

\section{EoR Simulations}\label{sec:ambermocks}
\begin{figure*}
    \centering
    \includegraphics[width=\textwidth]{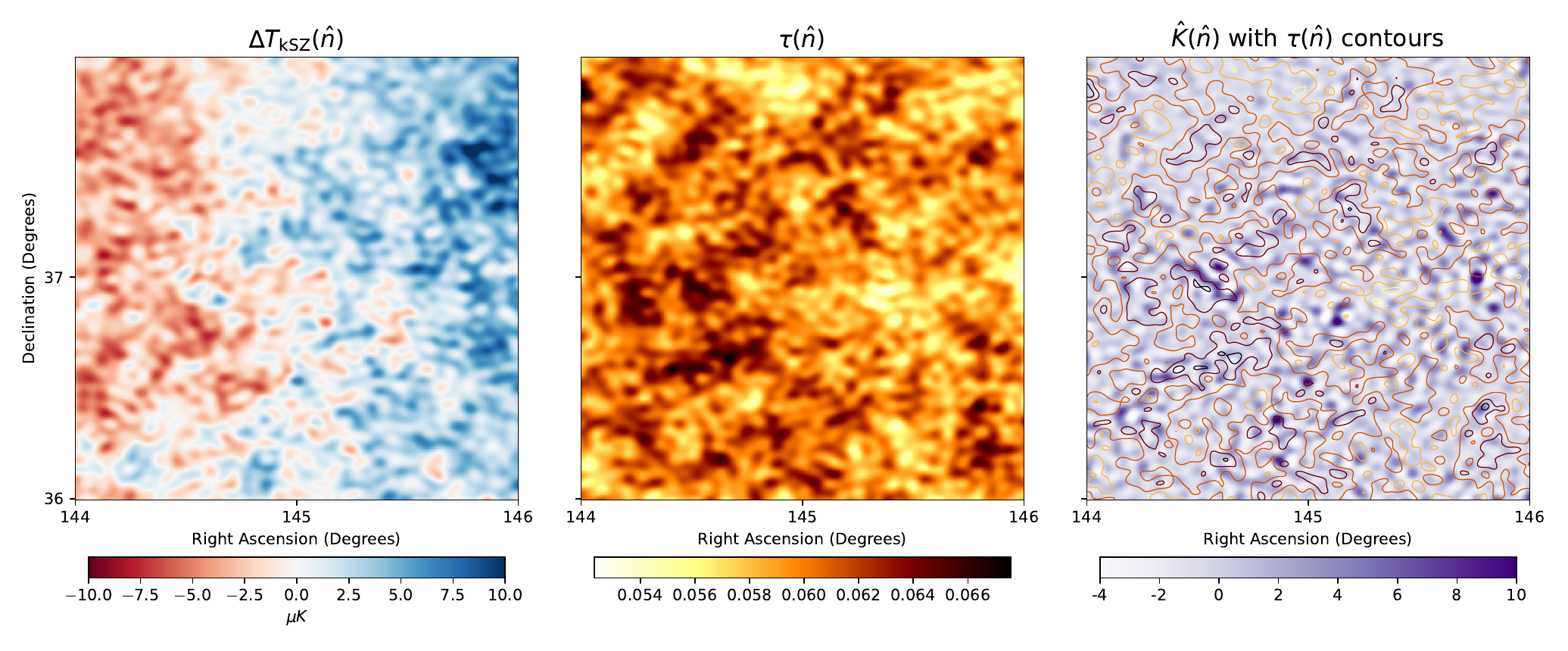}
    \caption{A 3-panel display of $2^{\circ} \times 2^{\circ}$ cutouts of the fiducial \texttt{AMBER} maps used in this work. \textit{Left:} The reionization kSZ map showing the changes to CMB temperature that the kSZ effect makes. \textit{Center:} The $\tau$ map showing the projected electron density fluctuations centered about $\bar{\tau} = 0.06$. We have added the low-z value of $\bar{\tau}_{z<5} = 0.029$ to the \texttt{AMBER} map to include the optical depth from redshifts below those simulated by \texttt{AMBER}. \textit{Right:} In purple, the $\hat{K}(\hat{\mathbf{n}})$ map reconstructed using the kSZ map on the left and overlaid with colored $\tau$ contours from the center map to display the correlations between the two.} 
    \label{fig:map3panel}
\end{figure*}

To obtain a realistic signal for both patchy $\tau$ and kSZ during the EoR, we use the semi-numerical Abundance Matching Box for the Epoch of Reionization (\texttt{AMBER}) code \citep{AMBER2022,Chen2023}. \texttt{AMBER} is used to make full-sky Healpix simulations of different fields in the Universe from the EoR, such as CMB lensing, patchy $\tau$, and patchy kSZ. This code allows us to specify the reionization history, making it ideal for parameter-space studies and for inferring when reionization occurred.

There are five reionization parameters that can be varied within \texttt{AMBER} to obtain different reionization fields: the redshift midpoint $z_{\text{mid}}$, the 5\% -- 95\% ionized redshift duration $\Delta z$, the redshift asymmetry $A_z$, the minimum halo mass $M_{\text{min}}$, and the radiation mean-free path $l_{\text{mfp}}$. See \citet{AMBER2022} for more details about these parameters. We simulate maps with the same settings presented in \citet{Chen2023}, for which these parameters were varied individually about their central fiducial values [$z_{\text{mid}} = 8$, $\Delta z = 4$, $A_z = 3$, $M_{\text{min}} = 10^8 M_{\odot}$, $l_{\text{mfp}} = 3$ Mpc/h], creating 11 different reionization histories spanning $5 < z < 15$. These simulations incorporate a flat $\Lambda$CDM set of parameters that is consistent with current observational constraints.

We generate these simulations as 2 Gpc/h boxes with $2048^3$ cells and $2048^3$ particles. This is a large enough box size with high enough resolution to capture both the large-scale fluctuations in the velocity field and the small-scale fluctuations in the $\tau$ field that we need for this study \citep{Chen2023}. In particular, we are interested in the sub-arcminute scales that would be probed with a high-resolution CMB survey. To create the maps, we set the Healpix resolution parameter $N_\mathrm{side}=32768$, corresponding to a pixel scale of roughly 6~arcsec. We generate the maps at this extremely high resolution in order to minimize the smoothing and aliasing that artificially change the power spectrum at the smallest scales due to the finite resolution of the simulation. After deconvolving the pixel window functions for $N_\mathrm{side} = 32768$, we decrease the resolution of the maps to $N_\mathrm{side}=8192$, or 24 arcsec, to speed up the subsequent analysis. We find that this method yields kSZ power spectra that agree with analytical expectations to within a few percent out to small scales of $l \approx 16000$. 

We use the simulated kSZ maps to reconstruct $\hat{K}$ maps as in Section \ref{subsec:Krecon} before cross-correlating them with their corresponding $\tau$ maps. We apply the same reconstruction pipeline software as that recently used on ACT data and simulations by \citet{maccrann2024}, which utilizes the public reconstruction code \texttt{falafel} among other standard packages. Using this pipeline we also calculate and subtract the $N^0$ bias on the reconstructed $C_L^{\hat{K}\hat{K}}$ power spectrum. This bias is a result of using a quadratic estimator: Due to the presence of Gaussian fluctuations in the kSZ simulations, $N^0$ arises as an additive term in $C_L^{\hat{K}\hat{K}}$. Following standard reconstruction methods, we calculate $N^0$ analytically using the power spectrum estimated directly from the realization of the simulation. Because there is only kSZ power in these simulations, this term is significantly smaller, by roughly an order of magnitude, than the nominal reconstructed noise for the surveys we consider. We then subtract this $N^0$ term from $C_L^{\hat{K}\hat{K}}$ to get an estimate of $C_L^{KK}$.

In this study, we do not reconstruct $\hat{\tau}$ from CMB maps as described in Section \ref{subsec:taunorm}, because $\tau$ maps are raw products of our simulation process. However, our forecasts for the reconstruction noise for each field are based on $\hat{\tau}$ reconstruction from data, ensuring that we treat the $\tau$ and $K$ sensitivity on equal footing. 

\begin{figure*}
    \centering
    \includegraphics[width=\textwidth]{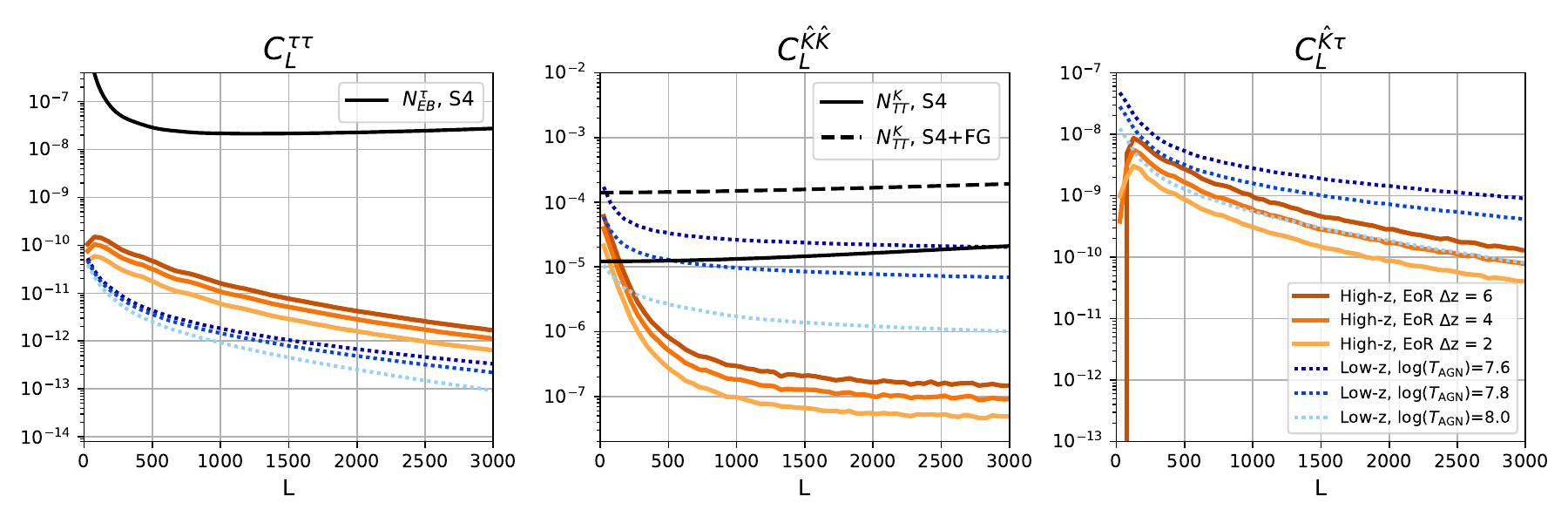}
    \caption{Comparison of the $\tau$, reconstructed $\hat{K}$, and $\hat{K}\times\tau$ signal power spectra from the EoR (\texttt{AMBER}) and low-redshifts (\texttt{Agora}) alongside their effective noise curves at $0 < L < 3000$. The orange signal curves are power spectra from three different \texttt{AMBER} models shown by the different shades in the legend. The black noise curves represent quadratic reconstructions of these fields, assuming white noise levels and beam for CMB-S4 temperature and polarization maps. The ``S4+FG" noise curve represents the post-ILC noise forecast for CMB-S4, which includes CMB foreground power. The blue dashed curves are the power spectra of the three \texttt{Agora} simulations (discussed in Section \ref{sec:agora}), demonstrating that they contribute roughly the same amount to the total as the EoR.}
    \label{fig:signalnoisecurves}
\end{figure*}

\begin{figure}
    \includegraphics[width=0.5\textwidth]{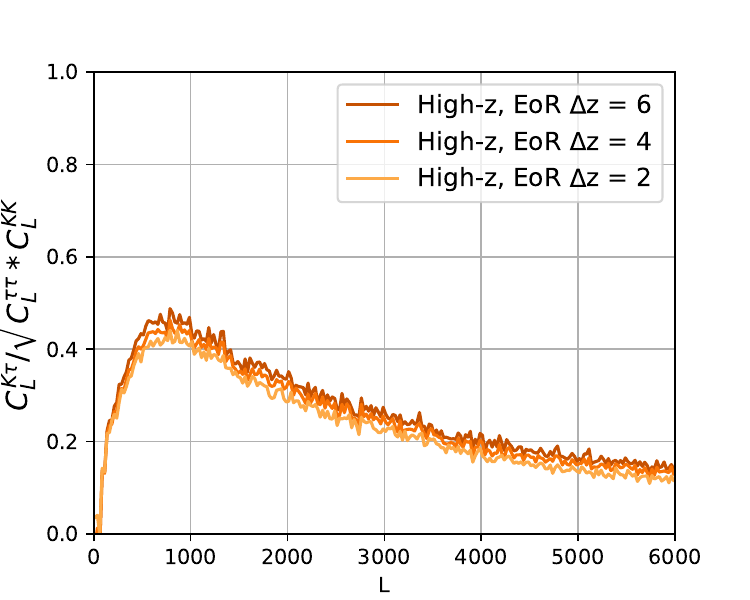}
    \caption{The dimensionless cross-correlation coefficient demonstrating the strength of the correlation between the fiducial $\hat{K}$ and $\tau$ \texttt{AMBER} fields up to $L = 6000$. This correlation peaks from $500 < L < 1000$ at close to 50\%. This is not a perfect correlation because the $\tau$ and $\hat{K}$ fields trace the electron density fields with different weights.}
    \label{fig:crosscorr}
\end{figure}

Fig. \ref{fig:map3panel} displays the set of \texttt{AMBER} maps that represent the fiducial parameters in this study, and the signal curves in Fig. \ref{fig:signalnoisecurves} represent the auto- and cross-power spectra of the $\tau$ and $\hat{K}$ fields. For the first time ever we demonstrate that the $\hat{K}$ and $\tau$ fields are correlated in the rightmost panel of Fig. \ref{fig:signalnoisecurves}. Not only are they non-zero, but their amplitudes are also appreciable in comparison to their constituent maps' spectra, meaning that a detection of this signal could be roughly as possible as detecting the two autospectra in the future. Figure \ref{fig:map3panel} shows a small cutout of the fiducial $\hat{K}$ map with $\tau$ overlays, showing that some correlation is visible by eye. By evaluating the cross-correlation statistic $\frac{C_L^{\hat{K}\tau}}{\sqrt{C_L^{\hat{K}\hat{K}} C_L^{\tau\tau}}}$, we find that these maps are close to 50\% correlated from $500 < L < 1000$ (see Fig. \ref{fig:crosscorr}). This imperfect correlation is likely the result of the $\tau$ and $K$ fields tracing differently-weighted versions of the electron density field, together with some effective $K$ reconstruction noise due to the nonzero power in the kSZ simulations.

\section{Sensitivity}\label{sec:noise}
\begin{figure}
    \includegraphics[width=0.5\textwidth]{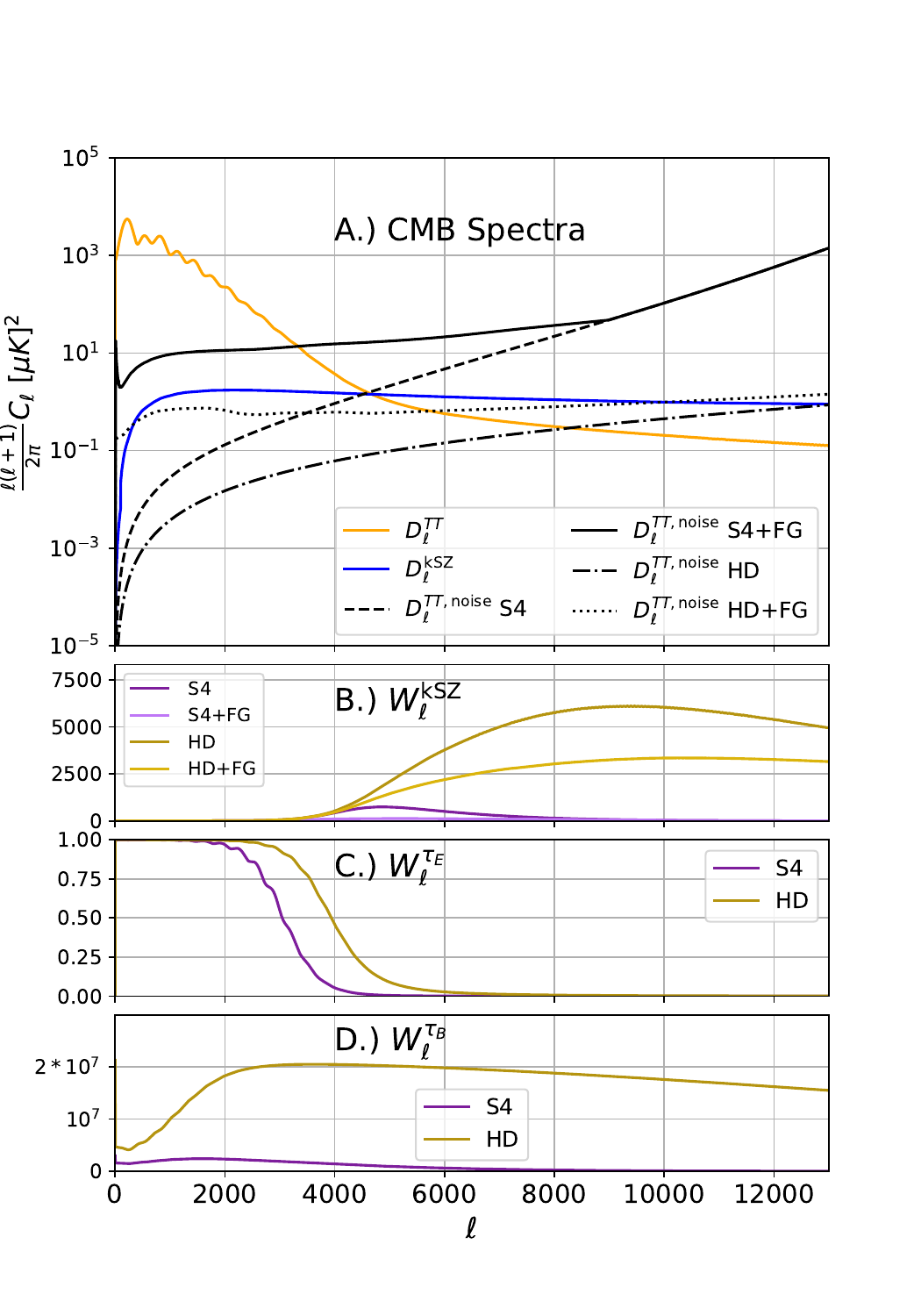}\label{fig:filters}
    \caption{A demonstration of all the components used to forecast the signal-to-noise ratios of $C_L^{KK}$, $C_L^{\tau\tau}$, and $C_L^{K\tau}$. A.) A plot of the lensed primary CMB temperature power spectrum, the fiducial kSZ power spectrum, and the four temperature noise curves for the different surveys used in this work. B.) The effective kSZ filters, $W_l^\mathrm{kSZ} \equiv \sqrt{C_l^{TT,\mathrm{kSZ}}} / C_l^{TT\mathrm{,t}}$ used to reconstruct $\hat{K}$ in this work, demonstrating the sensitivities of the different survey configurations to the kSZ signal. C.) The filters on the CMB E-mode polarization, $W_l^{\tau_{E}} \equiv C_l^{EE} / C_l^{EE\mathrm{,t}}$ for S4 and HD $\tau$ reconstruction calculations. D.) The filters on the CMB B-mode polarization, $W_l^{\tau_{B}} \equiv 1 / C_l^{BB\mathrm{,t}}$ for S4 and HD $\tau$ reconstruction calculations. Together these curves demonstrate the higher sensitivity that CMB-HD would have compared to CMB-S4. For kSZ science specifically, CMB-HD has much better sensitivity at much smaller scales than CMB-S4, even when foreground power is considered.}
\end{figure}

In this section, we lay out the steps necessary for making forecasts of this cross-correlation with CMB-S4 and CMB-HD \citep{sehgal2020, aiola2022}, two future CMB surveys which will achieve higher resolution and lower noise than current surveys. These properties make them especially suitable to detect signals that have small-scale effects on the primary CMB. We show the components of these forecasts in Figure \ref{fig:filters} and describe each one in the steps below. 

The first step in forecasting the sensitivities of CMB-S4 and CMB-HD to $C_L^{K\tau}$ is obtaining primary CMB theory and noise spectra. These curves are necessary for calculating the filters that enter the reconstruction calculations laid out in Section \ref{sec:reconstruction}. For this study, we use lensed CMB primary TT, EE, and BB power spectra ($C_l^{TT},\ C_l^{EE},\ C_l^{BB}$) up to $l_\mathrm{CMB} \sim 13000$ (see the top panel of Figure \ref{fig:filters}) such that we gain as much signal-to-noise as possible without pushing the limits of our simulations. Based on recent forecasts made with these surveys, we choose to effectively ``de-lens" $C_l^{BB}$ by applying a multiplicative factor, $A_\mathrm{lens}$, that represents the delensing ability of each given survey. For CMB-S4, we assume $C_l^{BB}$ can be delensed by 85\%, making $A_\mathrm{lens} = 0.15$ as done in \citet{Hotinli2022}. For CMB-HD, we set $A_\mathrm{lens} = 0.085$ based on \citet{MacInnis2024}. We assume for simplicity that the delensing and screening reconstruction processes are independent, through in reality they are connected \citep{namikawa2021_bmodes,Bianchini2023, Jain2024}. We also must obtain a theoretical kSZ temperature power spectrum for filtering purposes. Here we utilize the fiducial theory $C_l^{TT,\mathrm{kSZ}}$ from \texttt{AMBER}. These theory spectra enter the calculations of the reconstruction noise, $N_{XY}^{\Gamma}$, through $C_{l}^{XX}$ in the denominator of the filtering function, Equation \ref{eq:gencapf}.

The next step in calculating forecasts is to construct the primary CMB noise powers, $C_l^{XX\mathrm{,\ noise}}$, using the CMB-S4 and CMB-HD noise levels and beams; these CMB noise spectra enter the calculations of the reconstruction noise via Equation \ref{eq:gencapf}. For CMB-S4, we combine the 90 and 150 GHz channel estimates from \citet{Raghunathan2023} to obtain roughly $2\ \mu$K arcmin temperature white noise and a $\sim 1.8$ arcmin beam; for CMB-HD, we combine the 90 and 150 GHz channel estimates from \citet{MacInnis2024} to obtain $\sim 0.53\ \mu$K arcmin temperature white noise with a $\sim 0.35$ arcmin beam. We multiply these temperature noise levels by $\sqrt{2}$ for our polarization white noise levels. To account for the effects of CMB temperature foreground power, we make separate forecasts using the post-ILC $C_l^{TT\mathrm{,\ noise}}$ for S4 \citep{Raghunathan2023} and HD \citep{MacInnis2024}. For S4, we extrapolate the original curve from \citet{Raghunathan2023} out to $l=13000$ and make it equal to the original S4 curve where the beam begins to dominate (around $l = 9000$). We only consider these foregrounds for $C_l^{TT\mathrm{,\ noise}}$ because the polarization channels are much less plagued by foregrounds than temperature (as discussed previously in Section \ref{sec:intro}). Because of these two additional choices, we have four sets of forecasts: CMB-S4 (S4), CMB-S4 post-ILC (S4+FG), CMB-HD, and CMB-HD post-ILC (HD+FG), where the S4 and HD analyses do not  consider any foreground power.

Finally, we calculate the reconstruction noise spectra $N_{XY}^{\Gamma}$ according to Equation \ref{eq:gennorm} for $\Gamma=K$ and $\Gamma=\tau$ using the public code \texttt{tempura} (recall that $N^{\Gamma}_{XY}(\mathbf{L}) = A^{\Gamma}_{XY}(\mathbf{L})$ due to ideal filtering). These are considered the bandpower uncertainties per mode in reconstructed $\hat{K}$ and $\hat{\tau}$ maps. To forecast $\sigma (\hat{C}_L^{K\tau})$, we adopt the formula derived in \citet{knoxOG1995} to this work:
\begin{multline}\label{eq:knox}
    \sigma(\hat{C}_L^{K\tau})^2 = \\
    \lambda \left[ (C_L^{K\tau})^2 + 
    \left(C_L^{KK} + N_{TT}^{K}(L)\right)\left(C_L^{\tau\tau} + N_{EB}^{\tau}(L)\right) \right],
\end{multline}
with $\lambda = \frac{1}{(2L+1)\ f_{\mathrm{sky}}\ \Delta L}$, $f_{\mathrm{sky}}$ being the sky fraction of the given experiment \citep[here we set $f_\mathrm{sky} = 0.5$ based on the galaxy mask for CMB-S4 Wide from][]{Raghunathan2023} and $\Delta L$ being the chosen bin size (set to 500 for plots and 1 for SNR calculations). We use the \texttt{AMBER} fiducial $C_L^{\hat{K}\tau}$ result from Figure \ref{fig:signalnoisecurves} for this calculation. Note that this equation only applies explicitly to the calculation of $\sigma(\hat{C}_L^{K\tau})^2$: we also forecast $\sigma(\hat{C}_L^{KK})^2$ and $\sigma(\hat{C}_L^{\tau\tau})^2$ using the auto-power spectrum version of this Knox formula with each set of respective $C_L^{\Gamma\Gamma}$ and $N_{XY}^{\Gamma}$. For our nominal forecasts we use CMB temperature to reconstruct the $\hat{K}$ signal and noise, but use polarization to reconstruct the $\tau$ noise (recall we do not reconstruct $\hat{\tau}$ here). This is due to the fact that if we instead used CMB temperature to forecast both $K$ and $\tau$ signals and noises, we would obtain a lower uncertainty but then be sensitive to the highly-dominant CMB temperature foreground trispectrum that was discovered in both \citet{raghunathan2024} and \citet{maccrann2024} (as discussed in Section \ref{sec:intro}).

\section{Low-Redshift Simulations}\label{sec:agora}
As mentioned earlier, the post-EoR Universe contributes a significant amount of signal to both $K$ and $\tau$ such that we expect any cross-correlation between the two fields to contain contributions from both epochs. The kSZ signal from gaseous halos has been measured many times since \citet{Hand2012}, while the screening effect from similar halos was recently constrained at $\sim 2\sigma$ in \citet{coulton2025}. Under reasonable modeling assumptions, the low-redshift kSZ power spectrum has a comparable amplitude and shape to that of the EoR kSZ spectrum, where the low-redshift $\tau$ power spectrum is expected to have roughly an order of magnitude lower amplitude than its EoR counterpart. \citet{SmithFerraro2017} predicted that the EoR $C_L^{KK}$ will dominate over the signal from lower redshifts at the largest scales ($L \lesssim 300$), but the small scale comparison is not yet clear. We therefore investigate the low-redshift contributions to $C_L^{K\tau}$.

To estimate the low-redshift contribution to this cross-correlation, we use the \texttt{Agora} simulations from \citet{Omori2024}. This suite of millimeter sky simulations contains several relevant components to CMB science, which were made by pasting gas profiles halos from the BAHAMAS simulation suite \citep{mccarthy2016} onto the \textsl{MultiDark Planck 2} dark matter N-body simulation halos \citep{behroozi2013,klypin2016}. Three options for the pasted gas models were implemented in \texttt{Agora}, each made with a unique choice for the strength of AGN feedback within halos, as represented by the log$(T_\mathrm{AGN})$ parameter. Here, we analyze the $\tau (0 < z < 3)$ and $\Theta_{\mathrm{kSZ}}(0 < z < 3)$ maps with gas profiles assuming log$(T_\mathrm{AGN})=\ $7.6, 7.8, and 8.0. Though this range of log$(T_\mathrm{AGN})$ has been found to not fully represent the parameter space allowed by current data, it does encompass an appreciable fraction of the most likely range \citep{bigwood2024}. 

In order to make a realistic estimate of the post-EoR kSZ signal, we mask out the tSZ clusters in \texttt{Agora} that will be detectable by CMB-S4 before reconstructing the $\hat{K}$ field. In accordance with \citet{Raghunathan2022} and the \texttt{Agora} redshift range of $0 < z < 3$, we mask the $\sim$ 80,000 most massive halos from the \texttt{Agora} kSZ maps. This corresponds to a rough mass limit of $M_{500c} \approx 10^{14} M_\odot$, though we specifically use the Gaussian S4-Wide $M_{500c}(z)$ curve from  Figure 5 panel C in \citet{Raghunathan2022}. We then process these masked kSZ maps in exactly the same manner as the same fields from the \texttt{AMBER} simulations: We reconstruct $\hat{K}(\hat{\mathbf{n}})$ and cross-correlate the result with the corresponding $\tau(\hat{\mathbf{n}})$ fields. Figure \ref{fig:signalnoisecurves} shows the \texttt{Agora} $C_L^{\tau\tau}$, $C_L^{\hat{K}\hat{K}}$, and $C_L^{\hat{K}\tau}$ for the three different AGN models. For all three power spectra, we find the low-redshift component from \texttt{Agora} to have a similar shape to the EoR component from \texttt{AMBER}. Furthermore, $C_L^{\hat{K}\tau}$ seems to have comparable amplitudes from the EoR and post-EoR at the scales studied here. This is sensible when noting that reionization dominates the patchy $\tau$ signal for all models considered, and that the post-EoR dominates our reconstructed $\hat{K}$ signal at $L \geq 200$. We note that without masking these resolved tSZ clusters, the \texttt{Agora} $C_L^{\hat{K}\hat{K}}$ and $C_L^{\hat{K}\tau}$ signals would be overestimated by up to an order of magnitude.

\section{Forecasts}\label{sec:forecasts}
\begin{figure}[tp]
    \hspace{-8mm}
    \includegraphics[width=0.5\textwidth]{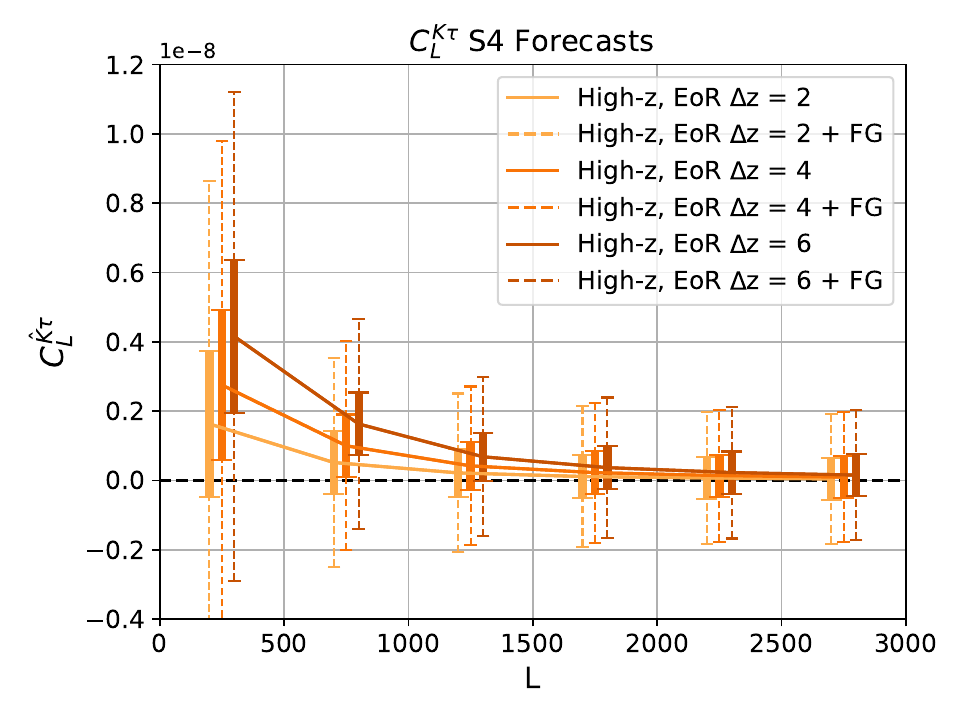}
    \caption{Forecasts of the CMB-S4 error bars on different cross-power spectra between the $\tau$ and $\hat{K}$ fields corresponding to different EoR redshift durations. The solid, thick error bars represent the forecasts that only consider the CMB-S4 white noise level and beam, and the thin dashed error bars also contain the foreground power with the post-ILC noise level from CMB-S4. This demonstrates that residual foreground power has a significant impact on the $C_L^{K\tau}$ SNR for CMB-S4.}\label{fig:s4forecasts}
\end{figure}

In this section we present the signal-to-noise forecasts on the cross-correlation between $\hat{K}$ and $\tau$ for CMB-S4 and CMB-HD. When we cross-correlate the $\tau$ and $\hat{K}$ signals from \texttt{AMBER} and use the forecasted cross-power spectrum uncertainties using Equation \ref{eq:knox}, we find that the fiducial EoR signal should be marginally detectable by CMB-S4 at $\SFsnr\sigma$ (see Fig. \ref{fig:s4forecasts} and Table \ref{tab:SNRtablefid}). However, when foreground power is considered in the temperature maps used to reconstruct $\hat{K}(\hat{\mathbf{n}})$, this forecast drops to $\SFPILCsnr\sigma$. For the case of CMB-HD, both types of forecasts look much stronger at $\HDsnr\sigma$ and $\HDPILCsnr\sigma$. Note that for CMB-S4, $\mathrm{SNR}(K\tau)$ is larger than $\mathrm{SNR}(\tau\tau)$ when ignoring foreground power in the temperature maps used to reconstruct $\hat{K}(\hat{\mathbf{n}})$, but this trend reverses when considering the temperature foregrounds. In contrast, $\mathrm{SNR}(K\tau)$ is larger than $\mathrm{SNR}(\tau\tau)$ in both cases for CMB-HD. Nevertheless, there is significant theoretical uncertainty on this cross-power spectrum: Figure \ref{fig:signalnoisecurves} demonstrates that the amplitude of the signal depends on EoR parameters such as the duration of reionization, which we investigate in the next section. It is thus possible that, in some reionization scenarios, this signal will be measured at higher significance than reported in these results.

\begin{figure}[tp]
    \hspace{-8mm}    
    \includegraphics[width=0.5\textwidth]{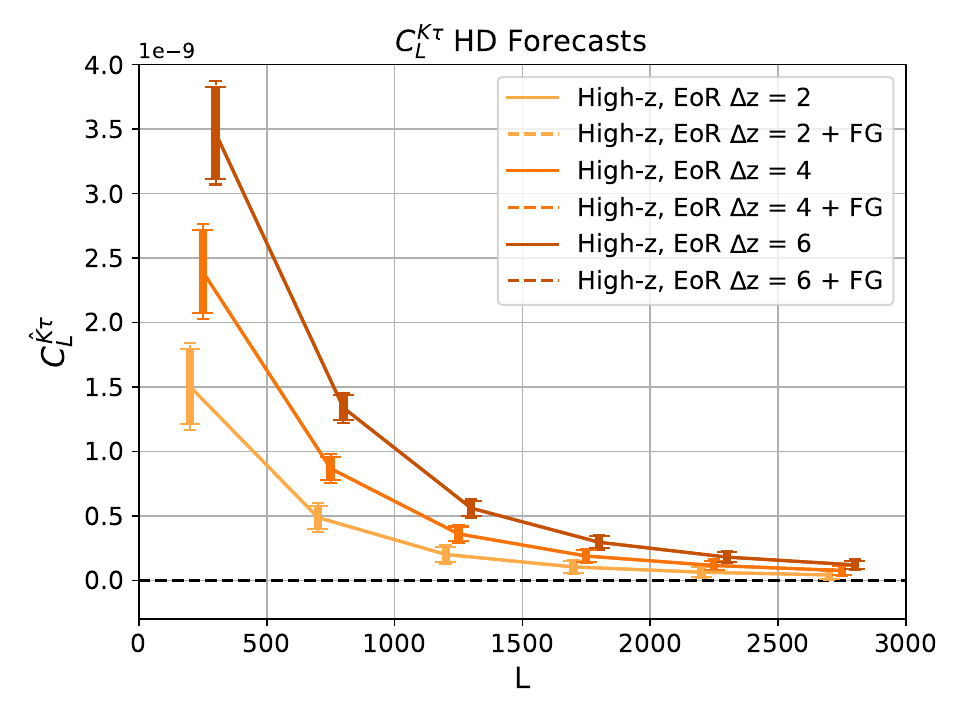}
    \caption{Forecasts of the CMB-HD error bars on different cross-power spectra between the $\tau$ and $\hat{K}$ fields corresponding to different EoR redshift durations. Foreground power is expected to have a much smaller impact on the size of these errors than for CMB-S4. Compared to CMB-S4 in Figure \ref{fig:s4forecasts}, both types of CMB-HD errors are much smaller and provide better SNR due to the reduced noise level, smaller beam, and more efficient delensing from CMB-HD.}\label{fig:hdforecasts}
\end{figure}
\begin{table}[h]
    \centering
    \begin{tabular}{|c|c|c|c|}
         \hline
             & \multicolumn{3}{c|}{\textbf{Signal-to-Noise Ratio}} \\ \hline
         \textbf{Survey} & $\mathbf{KK}$ & $\mathbf{\tau_{EB}\ \tau_{EB}}$ & $\mathbf{K \times \tau_{EB}}$ \\
         \Xhline{2\arrayrulewidth}
         CMB-S4 & 140 & 0.49 & \SFsnr \\
         \hline
         CMB-S4 with Foregrounds & 12 & 0.49 & \SFPILCsnr \\
         \hline
         CMB-HD & 2700 & 2.6 & \HDsnr \\
         \hline
         CMB-HD with Foregrounds & 1600 & 2.6 & \HDPILCsnr \\
         \hline
    \end{tabular}
    \caption{The signal-to-noise forecasts for fiducial ($\Delta z = 4$) $C_L^{KK}$, $C_L^{\tau_{EB}\tau_{EB}}$, and $C_L^{K\tau_{EB}}$ for CMB-S4 and CMB-HD with and without extragalactic CMB temperature foreground power (tSZ, CIB, and radio galaxies) and with $A_\mathrm{lens}=0.15, 0.085$ for S4 and HD, respectively. No foreground power is considered for $\tau_{EB}$ because polarization channels are much less contaminated by foregrounds than the temperature channel. While constraining $C_L^{KK}$ looks promising with both surveys, foregrounds have significant impacts on the noise. These impacts carry through to the cross-correlation between $K$ and $\tau$, making a detection unlikely with CMB-S4. However, CMB-HD will make a detection even with foregrounds included.}\label{tab:SNRtablefid}
\end{table}
We note that these results are constraints placed just on the amplitudes of the EoR components of $C_L^{K\tau}$, which is different from constraining this power spectrum from all redshifts. As depicted in the right panel of Figure \ref{fig:signalnoisecurves}, the low- and high-redshift components contribute a roughly equal amount of signal to the total cross-power spectrum, depending on the modeling choices assumed for both components. It is likely that by the time CMB-S4 and CMB-HD have the data to perform this analysis, we will have a better understanding of the distribution of ionized gas around the relevant halos due to, e.g., velocity-weighted stacking of the relevant objects \citep[e.g., ][]{schaan2021} or projected-fields measurements \citep[e.g., ][]{Hill2016, bolliet2023} at these low redshifts. Qualitatively, \citet{bolliet2023} forecast that CMB-S4 combined with a current galaxy survey like unWISE, using the projected-fields estimator, will constrain $C_L^{K \delta_g}$ to percent-level precision at low redshifts. \citet{roy2022probing} forecast that $C_L^{\tau g}$ will be measured at better than 10\% precision by CMB-S4 and Rubin. It follows that the low-redshift $C_L^{K \tau}$ will be measured to within about 10\% with CMB-S4 and a future galaxy survey. If we were to apply this uncertainty to the fiducial \texttt{Agora} model used here (the middle blue line in Figure \ref{fig:signalnoisecurves}) as a prior, it would be a few times smaller than the spread in the reionization models from \texttt{AMBER}, making it a subdominant source of uncertainty. In that case, the low-redshift components could be subtracted from $C_L^{K\tau}$ measurements with minimal residual uncertainty, yielding constraints that correspond more closely to the forecasts presented here.

A bias that will persist in this cross-correlation comes from CMB lensing, as found in the projected-fields kSZ studies \citep[e.g., ][]{Hill2016}. This is due to the similar methods of measurement the two effects have; specifically, that the lensing, screening, and squared kSZ response functions in Table 
\ref{tab:filters} are not decoupled from each other. Mode coupling from CMB lensing will thus lead to biased results in $C_L^{K\tau}$. Exploiting this synergy, a bias-hardening step in the measurement of $C_L^{K\tau}$ can eliminate the lensing bias at leading order for the cost of a small amount of noise \citep{namikawa2013}. To quantify the potential effects of lensing bias hardening on the uncertainties in $C_L^{KK}$ and  $C_L^{K\tau}$, we use the same pipeline described in \citet{maccrann2024} to compare the bias-hardened reconstruction noise, $N_{TT}^{K}$, to the noise calculated without considering lensing. We find that this bias hardening would cause a $\sim 20\%$ increase in $N_{TT}^{K}$  for the CMB-S4 cases and a $\sim 7\%$ increase for the CMB-HD cases, meaning that the cross spectrum sensitivity would degrade by 10\% and 4\% respectively. Thus, lensing hardening would not have a detrimental effect on the signal-to-noise ratio. Furthermore, the delensing we applied to the $B$ maps would remove most lensing signal from the $\tau$ maps, which we forecast based on the $EB$ coupling. Further lensing bias hardening could also be applied to these $\tau$ maps, and would yield negligible noise penalty \citep{roy2022probing}.

To avoid the likely dominant contribution from extragalactic temperature foregrounds, namely their flux trispectra, we conservatively focused on the polarization estimator for the $\tau$ forecasts. If we instead combine the polarization and temperature $\tau$ estimators, $\mathrm{SNR}(\tau\tau$) increases by $\sim 5$ times and $\mathrm{SNR}(K\tau)$ increases by $\sim 25-30$\%. This is notable for $C_L^{\tau\tau}$ in that it may cross the detection threshold with CMB-HD data. For the other survey cases, however, the $C_L^{\tau\tau}$ and $C_L^{K\tau}$ signals will likely remain below the detection limit, at least in our fiducial reionization scenario.

We acknowledge that using $\tau_{EB}$ or $\tau_{EB+TT}$ will still produce foreground biases in $C_L^{K\tau}$ due to the correlation between CMB foregrounds, such as the CIB or tSZ, with the $\tau$ field (i.e., even if reconstructed with polarization). This situation is analogous to the foreground contamination in CMB lensing cross-correlations as described in \citet{vanengelen2014}, \citet{osborne2014}, and \citet{Sailer2020}. This will naturally be suppressed by the application of internal linear combination methods to the multi-frequency CMB maps, which will reduce the contribution from these signals at the two-point level; this is what we have assumed for our nominal results that include foreground contribution. However, some contribution from CIB-$\tau$ and tSZ-$\tau$ correlation will likely remain. To test the case that we would ensure full deprojection of either (nominal) CIB or tSZ, we additionally used the CMB-S4 ``CIB-free" and ``tSZ-free" curves from \citet{Raghunathan2023} to reconstruct $N_L^{KK}$ and recalculate SNRs for $C_L^{K\tau}$. These deprojections come at the cost of additional residual noise.  We found that the CIB-free case yields a 55\% lower SNR for $C_L^{K\tau}$ than the original minimum variance case, while the tSZ-free case reduces the SNR by 85\%. Therefore, $C_L^{K\tau}$ is highly sensitive to CIB and tSZ deprojection. We note that one could instead do a cross-ILC approach, which is demonstrated in \citet{Raghunathan2023}. This method involves cleaning the CIB from one leg and the tSZ from the other in the $K$ estimator, significantly improving the foreground biases while yielding comparable SNR to the CIB-free case.

\section{Theory Interpretation}\label{sec:theoryint}
\subsection{Parameter Dependencies} \label{subsec:paramdep}
\begin{figure*}
    \centering
    \includegraphics[width=\textwidth]{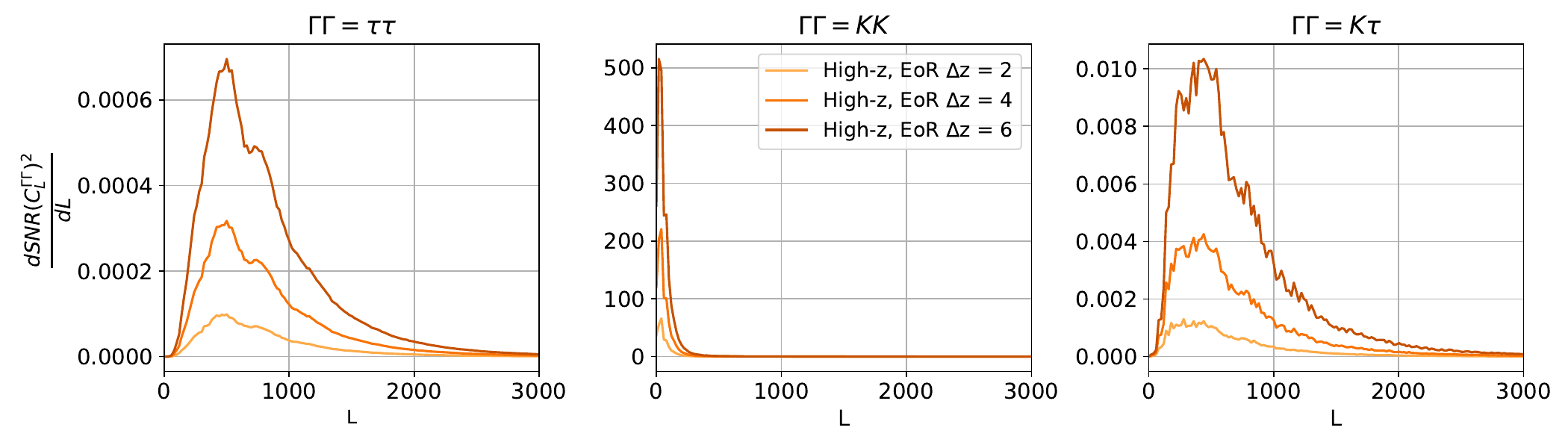}
    \caption{The contribution from each angular scale $L$ to the forecasted signal-to-noise ratios of $C_L^{\tau\tau}$, $C_L^{KK}$, and $C_L^{K\tau}$ for the case of CMB-S4 (neglecting foreground power). The SNRs of all three are dominated by large scales, though $C_L^{KK}$ is highly dominated by the very largest scales, $L<300$, due to its sensitivity to velocity fluctuations. In $C_L^{K\tau}$, we do not see this same strong dominance at $L<100$ because the constituent fields only have the $\tau$ fluctuations in common. This is also reflected in the power spectra in Figure \ref{fig:signalnoisecurves}. }
    \label{fig:dsnrdl}
\end{figure*}

Since we found that detecting $C_L^{K\tau}$ may be within reach of upcoming CMB surveys, here we investigate how such a measurement would impact our understanding of the reionization process. \citet{Chen2023} analyzed the dependence of the kSZ power spectrum on the various reionization parameters in the \texttt{AMBER} simulations, finding that the kSZ signal is most dependent on $\Delta z$ when considering the large range of allowed values for $\Delta z$ by current data. Forecasts show that $C_L^{KK}$ is degenerate in $\Delta z$ and $\bar{\tau}$ \citep{FerraroSmith2018}, though when used in combination with the kSZ power spectrum and the spatially-averaged optical depth $\bar{\tau}$ from \textsl{Planck}, this degeneracy is broken and tight constraints can be placed on both \citep{alvarez2021}. Here, we inspect the parameter dependence of the \texttt{AMBER} $C_L^{\tau\tau}$, $C_l^{TT,\mathrm{kSZ}}$, $C_L^{\hat{K}\hat{K}}$, and $C_L^{\hat{K}\tau}$ on each of the reionization parameters included in the simulations. To do this, we assume that near the fiducial model $C_L^{XY, \mathrm{fid}}$, each power spectrum $C_L^{XY}$ scales with the parameters according to
\begin{equation}\label{eq:parameterization}
    C_L^{XY} = C_L^{XY, \mathrm{fid}} \prod_{i=1}^{5}\left(\frac{p_i}{p_i^{\mathrm{fid}}}\right)^{s_i},
\end{equation}
for $p_i \in$ [$z_{\text{mid}}$, $\Delta z$, $A_z$, $M_{\text{min}}$, $l_{\text{mfp}}$] and their corresponding power-law indices $s_i$. As described in Section \ref{sec:ambermocks}, when generating our reionization simulations we vary these parameters individually about a central value such that we have three power spectra per parameter. We then compute the logarithmic derivative $\partial \mathrm{ln}(C_L^{XY})/\partial \mathrm{ln}(p_i)$ to isolate each index $s_i$, using centered finite differences with the \texttt{AMBER} mocks. This enables us to quantify the strength of dependence of each power spectrum on each parameter. To present these quantitatively, we show the results at a binned $L$ value that corresponds roughly to the peak in the $\frac{\mathrm{dSNR}^2}{\mathrm{d}L}$ curves for each $C_L$, indicating the parameter sensitivity in the region of the spectrum that will be measured. We show these curves in Figure \ref{fig:dsnrdl} and the results of this investigation in Table \ref{tab:param_powers}, where higher values mean stronger dependence. 

We find that when viewed as parameter derivatives, the spectra considered here are most sensitive to $z_\mathrm{mid}$. However, $z_\mathrm{mid}$ is relatively well-constrained with other methods \citep[e.g., Table 2 in][]{planck2018} compared to $\Delta z$ which still has a significant uncertainty \citep[e.g., ][]{raghunathan2024}. Nevertheless, this investigation indicates that an upper limit on or detection of $C_L^{K\tau}$ will be most informative on both $z_\mathrm{mid}$ and $\Delta z$ compared to the other reionization parameters. All the parameter sensitivities for the auto- and cross-power spectra of the $K$ and $\tau$ fields are shown in Table \ref{tab:param_powers}: $C_L^{KK}$ is generally more dependent on most parameters than the other three spectra. The one exception is that $C_L^{K\tau}$ is more sensitive to $A_z$, the asymmetry of reionization in redshift, than the three autospectra. Since $C_L^{K\tau}$ has significant sensitivity to both $\Delta z$ and $A_z$, it is a promising probe for the first half of reionization and would be complementary to probes that are sensitive to the second half, such as high-redshift quasar spectra \citep[e.g.,][]{Davies2018,Bosman2022}. Due to the low SNR forecasts for $C_L^{K \tau}$, we chose not to do a deeper Fisher analysis on these parameters and leave that to future work with data, or in tandem with other EoR observables.

We note that reionization parameters are not fully orthogonal to cosmological ones. Specifically, $C_L^{\tau\tau}$ and $C_L^{TT\mathrm{, kSZ}}$ are also sensitive to the clustering of matter, parameterized as $\sigma_8$, due to the fact that they trace the density and velocity fluctuations. \citet{Chen2023} studied the dependence of the kSZ autospectrum on cosmology using the same \texttt{AMBER} mocks and found that, roughly,  $C_L^{TT\mathrm{, kSZ}} \propto \sigma_8^{4.2}$. If we assume similar scalings for higher-point functions as for the kSZ power spectrum, and that the large-scale bulk-flow velocities scale linearly with $\sigma_8$, $C_L^{K \tau}$ only contains one extra power of $\delta_e$, so likely scales roughly as $\sigma_8^{5}$ -- $\sigma_8^{6}$. Factoring in the current uncertainty on $\sigma_8$ of $\sim$ 1\% \citep[e.g., ][]{planck2016, madhavacheril2024}, the predicted scatter in $C_L^{K \tau}$ from uncertain cosmology is $\sim 6$\%. We can compare this to the expectation for uncertain reionization parameters: using the current-best reionization parameter constraint \citep[$z_\mathrm{mid} = 8.14 \pm 0.61$;][]{planckhfi2020} together with the power-law scaling from Table \ref{tab:param_powers}, we find that $C_L^{K \tau}$ has $\sim12$\% scatter from $z_\mathrm{mid}$ uncertainty. This demonstrates that we expect cosmological uncertainties to affect $C_L^{K \tau}$ significantly less than the uncertainties in reionization parameters.

\begin{table}
    \centering
    \begin{tabular}{|c|c|c|c|c|c|}
        \hline
        & \multicolumn{5}{c|}{\textbf{Power Law Scaling}} \\
        \hline
        \textbf{Spectrum} & $z_{\text{mid}}$ & $\Delta z$ & $A_z$ & $M_{\text{min}}$ & $l_{\text{mfp}}$ \\
        \Xhline{2\arrayrulewidth}
        $C_L^{\tau\tau}$ & 2.2 & 0.93 & 0.11 & 0.003 & 0.54 \\
        $(450<L_\mathrm{ref}<550)$ & & & & & \\
        \hline
        $C_l^{TT,\mathrm{kSZ}}$ & 1.4 & 0.75 & 0.012 & 0.001 & 0.03 \\
        $(2950<l_\mathrm{ref}<3050)$ & & & & & \\
        \hline
        $C_L^{KK}$ & 2.7 & 1.1 & 0.12 & 0.0009 & 0.10 \\
        $(150<L_\mathrm{ref}<250)$ & & & & & \\
        \hline
        $C_L^{K\tau}$ & 1.6 & 1.1 & 0.26 & 0.0001 & 0.29 \\
        $(350<L_\mathrm{ref}<450)$ & & & & & \\
        \hline
    \end{tabular}
    
    \caption{Power law scalings from Equation \ref{eq:parameterization} for each parameter in the \texttt{AMBER} simulations and each power spectrum in this analysis, demonstrating their dependence on the parameters. The $L_\mathrm{ref}$ values indicate the bin for which these values were calculated, roughly based on the peaks in the signal-to-noise derivative with respect to $L$ for each power spectrum, shown in Figure \ref{fig:dsnrdl}. Each spectrum is most sensitive to $z_\mathrm{mid}$, with $\Delta z$ a close second for most.}
    \label{tab:param_powers}
\end{table}

\subsection{Source of the Signal}\label{subsec:bispectest}
\begin{figure}
    \includegraphics[width=\linewidth]{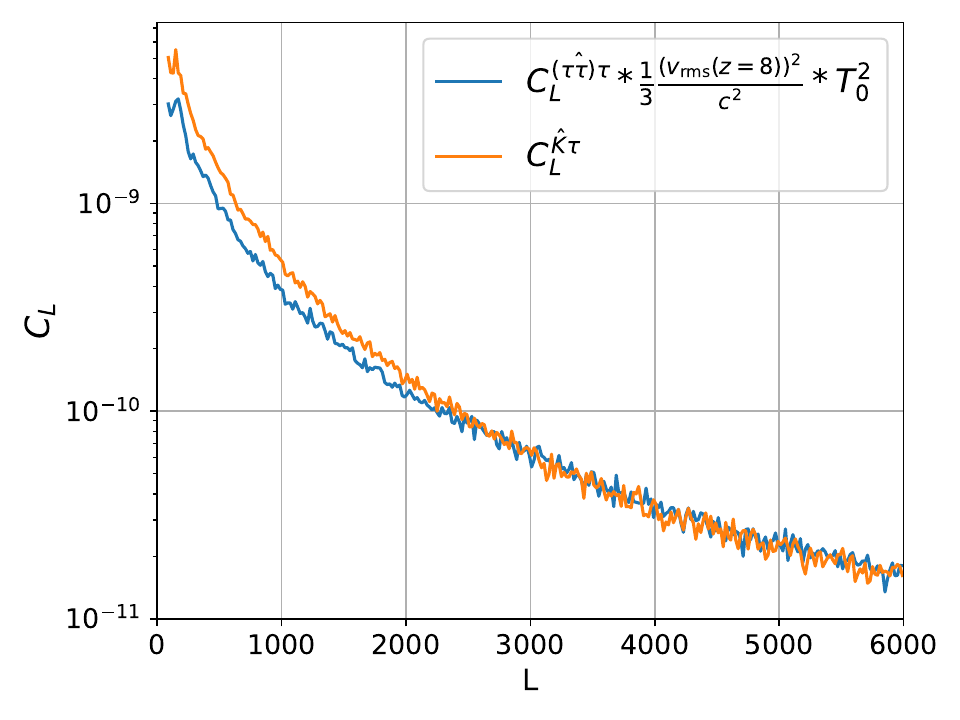}
\caption{Demonstration of the ansatz in Equation \ref{eq:bispec}, showing that our estimate of the theoretical $C_L^{K\tau}$ (orange) is dominated by the electron density bispectrum (blue), especially on small scales of $L>3000$. Even on large scales, the $\delta_e$ bispectrum scaled appropriately by $v^2_\mathrm{rms}$ contributes to over half of $C_L^{K\tau}$. This confirms that this cross-correlation almost solely probes the non-Gaussianity of electrons during the EoR.} \label{fig:taubispec}
\end{figure}

To further investigate the theory of $C_L^{K\tau}$ introduced in Section \ref{sec:formalism}, we perform a test to check what signals this cross-correlation is sourced by. Based on Equations \ref{eq:clkt}-\ref{eq:bispecdef}, it is reasonable to ask whether $C_L^{K\tau}$ is dominated by the electron density fluctuations scaled by $v^2_\mathrm{rms}$ because it depends on the hybrid bispectrum between the electron density and momentum fields. On small scales, velocities vary slowly across the sky, so within a local patch we can potentially approximate $p(\hat{\mathbf{n}}) \approx v(\hat{\mathbf{n}}) \left(1+\delta_e(\hat{\mathbf{n}})\right) \approx v_\mathrm{loc} \left(1+\delta_e(\hat{\mathbf{n}})\right)$ for the local value of the velocity. This is typically taken to be the dominant regime for projected-fields kSZ studies, as first assumed by \citet{Dore2004}. \citet{DeDeo2005} investigated the assumptions from \citet{Dore2004} in depth, and verified that this $\langle v v  \rangle \langle \delta \delta \delta \rangle$ term in the full $\langle p_{\hat{\mathbf{n}}} p_{\hat{\mathbf{n}}} \delta \rangle$ bispectrum, mirrored in Eq. \ref{eq:bispecdef}, is dominant on small scales. Though \citet{patki2023} recently found that some terms besides $\langle vv \rangle \langle \delta \delta \delta \rangle$ are stronger than \citet{DeDeo2005} expected, we seek to determine whether this term should contribute the large majority of the $C_L^{K\tau}$ signal at the scales we probe in this work. Here we investigate whether $C_L^{K\tau}$ from the EoR is also within this regime at the scales where it will be measured with upcoming surveys.

We test the approximation by passing the fiducial \texttt{AMBER} $\tau(\hat{\mathbf{n}})$ map through the $K$ reconstruction pipeline, obtaining a normalized $\sim \tau_\mathrm{small-scale}^2$ map in place of the usual $(\Theta_\mathrm{kSZ,\ small-scale}^\mathrm{filt})^2$ map. We then cross-correlate this reconstructed map with the original $\tau(\hat{\mathbf{n}})$ map. This directly yields the electron density bispectrum contribution to our overall $C_L^{K\tau}$ calculation, $B_{\delta_e \delta_e \delta_e}$. We make the ansatz that this allows us to estimate $B_{\delta p_{\hat{\mathbf{n}}} p_{\hat{\mathbf{n}}}}$ in $C_L^{K\tau}$ from Equation \ref{eq:triangle} via:
\begin{equation}\label{eq:bispec}
    B_{\delta_e p_{\hat{\mathbf{n}}} p_{\hat{\mathbf{n}}}} \approx \frac{1}{3} v_{\mathrm{rms}}^2 T_0^2 B_{\delta_e \delta_e \delta_e}.
\end{equation}
Because $C_L^{K\tau}$ contains information about $v_\mathrm{rms}$, we multiply this projected $B_{\delta_e \delta_e \delta_e}$ by $\frac{1}{3}\frac{v^2_\mathrm{rms}
}{c^2}T_0^2$ to make an apples-to-apples comparison of this bispectrum and the original $C_L^{K\tau}$. The $1/3$ scaling comes from the fact that we take only the radial component of the three-dimensional velocity field in this projection, and the scaling by the primary CMB temperature $T_0$ corrects for the fact that $\tau$ maps are unitless, while our kSZ maps are in $\mu$K. Given that the rms value of the bulk-flow velocities is not a strong function of redshift, for this test we replace it by a constant value at the fiducial midpoint of reionization in our fiducial \texttt{AMBER} mocks, $z = 8$. We use the Code for Anisotropies in the Microwave Background \citep[CAMB,][]{Lewis2000} to estimate $v^2_\mathrm{rms}(z=8) \approx 240$ km\,s$^{-1}$.

The result of this test is shown in Figure \ref{fig:taubispec} and demonstrates that these two bispectra are indeed in good agreement. On small scales, $(L>3000)$, they match as expected, and on larger scales this electron bispectrum constributes more than half of the total $C_L^{K\tau}$ signal. Therefore, our ansatz in Equation \ref{eq:bispec} is validated. This means that a measurement of $C_L^{K\tau}$ in the manner described in this work will directly probe $B_{\delta_e \delta_e \delta_e}$, something that will yield information solely about the skewness, or bispectrum, of the non-Gaussianity in the $\delta_e$ field during reionization. By contrast, the other higher-point functions we studied measure even-numbered statistics of the electron distribution during the EoR. Therefore, $C_L^{K\tau}$ contains unique, novel information. 

\section{Discussion}\label{sec:discussion}
We have introduced a new CMB estimator for studying reionization: the cross-correlation between reconstructed maps of patchy screening and the squared kinetic Sunyaev-Zel'dovich fluctuations, $C_L^{K\tau}$. These two types of reconstructed maps have been considered as separate probes of the reionization process; here, we investigated the potential gains from analyzing them together in cross-correlation. 

We found that the $C_L^{K\tau}$ signal is appreciable and positive because both trace the same bubbles of ionized gas during reionization. To determine the detectability of this signal, we considered the case in which both maps are reconstructed from the same CMB survey --- using the CMB temperature information for the $\hat{K}$ reconstruction, and, to avoid the leading source of foreground bias, using the CMB polarization information for estimating the patchy $\tau$ noise. We forecasted signal-to-noise ratios for the $\tau$ auto-power spectrum $C_L^{\tau\tau}$, the $K$ auto-power spectrum $C_L^{KK}$, and $C_L^{K\tau}$ using expected noise levels for the CMB-S4 and CMB-HD surveys. We found that the outlook for a detection of this signal is promising for future CMB experiments, specifically that CMB-S4 should make a $\SFPILCsnr-\SFsnr\sigma$ measurement and CMB-HD should make a $\HDPILCsnr-\HDsnr\sigma$ measurement, depending on the amount of power from other extragalactic foregrounds remaining in the kSZ maps after the multi-frequency ILC cleaning process is performed. Even considering the most pessimistic SNR, an upper limit on $C_L^{K\tau}$ will yield valuable information about the midpoint and duration of reionization via this new, independent method from the current best constraints. We quantified this by explicitly measuring how the auto- and cross-power spectra depend on the 5 reionization parameters varied in the \texttt{AMBER} mocks, finding significant sensitivity to these properties. Specifically, we found that $C_L^{K\tau}$ is a complementary way to measure $\Delta z$ and is uniquely sensitive to the skewness of reionization in redshift, $A_z$, making it a probe of the early phases of reionization. 

We also investigated how much the low-redshift gaseous halos surrounding galaxies contribute to $C_L^{K\tau}$ by using the \texttt{Agora} post-EoR simulations. We found that, after masking the brightest halos, these lower redshifts contribute a roughly equal amount of $C_L^{K\tau}$ signal as the EoR does. Even so, this low-redshift component will likely become less theoretically uncertain than it is now with the advent of new kSZ analyses at these low redshifts in the coming years. 

To avoid the most detrimental source of bias from other non-Gaussian extragalactic sources, which appear in the temperature four-point function, we conservatively focused on the polarization estimator for reconstructing the $\tau$ maps. However, it is possible that these terms could be brought under control using tools developed for CMB lensing analyses. If we instead use both the temperature and polarization estimators for the $\tau$ forecast, the signal-to-noise on $C_L^{K\tau}$ increases by $\sim 25-30$\% and that of $C_L^{\tau\tau}$ by a factor of $\sim 5$ times.

We found that $C_L^{K\tau}$ is mainly sourced by small-scale angular fluctuations in the optical depth and their cross-correlation with small-scale squared optical depth fluctuations. This makes it the high-redshift analog of the projected-fields method of measuring the kSZ signal. In this view, the $C_L^{K\tau}$ spectrum is directly probing the bispectrum, or spatial skewness, of the ionized bubbles during reionization. This makes it a distinct probe from both $C_L^{KK}$ and $C_L^{\tau\tau}$, which measure other aspects of the morphology of the reionization process.

In summary, we found that the new cross-correlation statistic we isolated will be a new and complementary way to constrain the epoch of reionization using upcoming CMB surveys. This further establishes the degree to which these surveys will be critical tools for uncovering the elusive nature of the earliest generations of stars and galaxies. 

\section*{Acknowledgments}
We would like to thank Victor Chan, Amanda MacInnis, Joel Meyers, Srini Raghunathan, and Neelima Sehgal for correspondence surrounding the experimental properties, including  foregrounds and delensing levels, that we used in our reconstruction filters (those shown in Figure \ref{fig:filters}). We thank Seth Cohen, Will Coulton, Simone Ferraro,   Simon Foreman, Phil Mauskopf, Yogesh Mehta, Srini Raghunathan,  and Shabbir Shaikh for their discussion and comments on this work. We also thank Nianyi Chen for correspondence about the \texttt{AMBER} simulations and Colin Hill for correspondence about the \texttt{Agora} simulations.

This work was supported by NASA ADAP grant 80NSSC24K0665. AvE and DK were additionally supported by NASA ADAP grants 80NSSC23K0747 and 80NSSC23K0464, and NSF AAG grant 588167.

\bibliography{BIB}

\end{document}